\documentclass{ieeeaccess}
\usepackage{cite}
\usepackage{amsmath,amssymb,amsfonts}
\usepackage{graphicx}
\usepackage{textcomp}
\usepackage{comment}
\usepackage{caption}

\usepackage{etoolbox}
\usepackage{enumitem}

\def\thevol{XX}
\def\myyear{XXXX}
\makeatletter
\patchcmd{\@evenfoot}{2016}{\myyear}{}{}
\patchcmd{\@oddfoot}{2016}{\myyear}{}{}
\makeatother

\def\BibTeX{{\rm B\kern-.05em{\sc i\kern-.025em b}\kern-.08em
    T\kern-.1667em\lower.7ex\hbox{E}\kern-.125emX}}

\usepackage{cite}
\usepackage{psfrag}
\usepackage{url}
\usepackage{stfloats}
\usepackage{amsmath}
\usepackage{array}
\usepackage{epsfig}
\usepackage{amssymb}
\usepackage{color}
\usepackage{subfig}
\usepackage{setspace}
\usepackage{caption}
\usepackage{booktabs}
\usepackage{ dsfont }
\usepackage{algpseudocode}

\usepackage[square,numbers,sort&compress]{natbib}
\usepackage{algorithm}  
\usepackage{amsmath,epsfig,amssymb,algpseudocode,amsthm,cite,url}
\usepackage{comment}

\usepackage{multirow}

\graphicspath{{"./pictures/"}}

\ifCLASSINFOpdf
\graphicspath{{./pictures/}}
\else
\usepackage{enumerate}
\usepackage{amsfonts,amssymb}
\fi

\begin{document}
\history{Date of publication xxxx 00, 0000, date of current version xxxx 00, 0000.}
\doi{xx.xxxx/ACCESS.2020.DOI}

\title{Coverage Evaluation for 5G Reduced Capability New Radio (NR-RedCap)}
\author{\uppercase{Saeedeh Moloudi}\authorrefmark{1}, 
\uppercase{Mohammad Mozaffari}\authorrefmark{2}, \uppercase{Sandeep Narayanan Kadan Veedu}\authorrefmark{1},  \uppercase{Kittipong Kittichokechai}\authorrefmark{1}, \uppercase{Y.-P. Eric Wang}\authorrefmark{2}, \uppercase{Johan Bergman}\authorrefmark{3}, and \uppercase{Andreas Höglund}\authorrefmark{1}\vspace{0.2cm}}
\address[1]{Ericsson Research, Sweden, (e-mails: \{saeedeh.moloudi, sandeep.narayanan.kadan.veedu, kittipong.kittichokechai, andreas.hoglund\}@ericsson.com)}
\address[2]{Ericsson Research, Silicon Valley, Santa Clara, 
CA, USA, (e-mails: \{mohammad.mozaffari, eric.yp.wang\}@ericsson.com)}
\address[3]{Ericsson Business Unit Networks, Sweden, (e-mail:  johan.bergman@ericsson.com)}

\corresp{Corresponding author: Saeedeh Moloudi (e-mail: saeedeh.moloudi@ericsson.com). 
}



\begin{abstract}
	The fifth generation (5G) wireless technology is primarily designed to address a wide range of use cases categorized into the enhanced mobile broadband (eMBB), ultra-reliable and low latency communication (URLLC), and massive machine-type communication (mMTC). Nevertheless, there are a few other use cases which are in-between these main use cases such as industrial wireless sensor networks, video surveillance, or wearables.  In order to efficiently serve such use cases, in Release 17, the 3rd generation partnership project (3GPP) introduced the reduced capability NR devices (NR-RedCap) with lower cost and complexity, smaller form factor and longer battery life compared to regular NR devices. However, one key potential consequence of device cost and complexity reduction is the coverage loss. In this paper, we provide a comprehensive evaluation of NR RedCap coverage for different physical channels and initial access messages to identify the channels/messages that are potentially coverage limiting for RedCap UEs. We perform the coverage evaluations for RedCap UEs operating in three different scenarios, namely Rural, Urban and Indoor with carrier frequencies 700 MHz, 2.6 GHz and 28 GHz, respectively. Our results confirm that for all the considered scenarios, the amounts of required coverage recovery for RedCap channels are  either less than 1 dB or can be compensated by considering smaller data rate targets for RedCap use cases.
	

\end{abstract}


\titlepgskip=-1pt

\maketitle

\section{Introduction}
\label{sec:intro}

The fifth generation (5G) wireless technology enables a wide range of services with different requirements in terms of data rates, latency, reliability, coverage, energy efficiency, and connection density. Specifically, the 5G new radio (NR) primarily supports enhanced mobile broadband (eMBB) and ultra-reliable and low-latency communication (URLLC) use cases \cite{dahlman20185g, ahmadi}. The 5G NR caters to flexibility, scalability, and efficiency with its unique features and capabilities. It supports a wide frequency range, large bandwidth (BW), flexible numerology, dynamic scheduling, and advanced beamforming features which make it suitable for enabling various use cases with stringent data rate and latency requirements \cite{dahlman20185g}. Meanwhile, massive machine-type communication (mMTC) is supported by the low-power wide-area network (LPWAN) solutions such as long term evolution for machine-type communications (LTE-M) and narrowband Internet-of-Things (NB-IoT) \cite{GSMAwhite, liberg2017cellular, NRCoex}. In addition, there are still several other use cases whose requirements are higher than LPWAN (i.e., LTE-M/NB-IoT) but lower than URLLC and eMBB \cite{SID_redcap}. In order to efficiently support such use cases which are in-between eMBB, URLLC, and mMTC, the 3rd generation partnership project (3GPP) has studied reduced capability NR devices (NR-RedCap) in Release 17 \cite{SID_redcap}. The RedCap  study item has been completed in December 2020 and is continued as a work item \cite{WID_redcap}.

The NR-RedCap user equipment (UE) is designed to have lower cost, lower complexity (e.g., reduced bandwidth and number of antennas), a longer battery life, and enable a smaller form factor than regular NR UEs. These devices support all frequency range 1 (FR1) and frequency range 2 (FR2) bands for both frequency division duplex (FDD) and time division duplex (TDD) operations. One drawback of complexity reduction in terms of device bandwidth reduction or number of Rx/Tx antenna reduction for RedCap UEs is coverage loss which varies for different physical channels. To compensate for the coverage loss, different coverage-recovery solutions can be considered depending on the coverage-limiting channels and the level of the needed coverage recovery.

Our aim in this paper is to investigate the impact of the complexity reduction on the coverage performance of RedCap UEs, identify the corresponding coverage-limiting channels, and evaluate the amount of coverage recovery needed for those channels. For that, we have considered the Rel-15 NR UEs as a reference UE, and compared the coverage performance of RedCap UEs to the reference UE performance for all NR physical channels used for the initial access, random access, as well as control and data channels for downlink (DL) and uplink (UL) transmissions.

To evaluate the coverage performance, we have followed two main steps: 1) performed link-level simulations (LLSs) to obtain the required SINR, considering performance targets such as block error rate (BLER) for the different physical channels;  2) used the LLS results and performed link-budget evaluation for both reference UE and RedCap UE. Finally, considering maximum isotropic loss (MIL) as a coverage-evaluation metric, we have identified the reference UE channel with the lowest MIL as the bottleneck channel, (i.e. the channel that is limiting Rel-15 coverage), and the corresponding MIL as a coverage threshold. Any RedCap channel with MIL smaller than the threshold is considered as coverage limiting channel and needs coverage recovery. Our results  show that for RedCap UEs operating
in FR1 bands, PUSCH and Msg3 
need approximately 3 dB and 0.8 dB coverage recovery, respectively. In FR2, the impact of complexity reduction is more considerable for DL channels. Based on our results, PDSCH and Msg4 require 3.4 dB and 0.5 dB coverage recovery, respectively. It should be noted that the required coverage recovery for data channels can be compensated by reducing the data rate targets.  Our results demonstrate key tradeoffs and guidelines needed for designing the NR-RedCap.

The rest of the paper is organized as follows. First, we provide a list of main abbreviations (see Table \ref{Abbr}) used throughout the paper. In Section II, we provide an overview of NR-RedCap UEs and their key features. In Section III, a detailed description and results of LLSs are presented. Subsequently, Section IV covers our link budget evaluations. Finally, the concluding remarks are provided in Section V.


\begin{table}[!t]
	
	\caption{\small List of abbreviations.}
	\resizebox{.95\columnwidth}{!}{
		\begin{tabular}{@{}ll@{}}
			\toprule
		
				Abbreviation &	Definition \\ \midrule
	
		BW &	Bandwidth \\
		BWP&Bandwidth part in NR\\
		CRC&	Cyclic redundancy check\\
		CORESET& Control resource set\\
		DCI&	Downlink control information\\
		DL&	Downlink \\
		DMRS&	Demodulation reference signal\\
		FDD&	Frequency division duplex\\
		FDRA&	Frequency domain resource allocation\\
		LLS&	Link-level simulation \\
		LPWAN&	Low-power wide-area network \\
		LTE-M& Long term evolution for machine-type communications\\
		MCS&	Modulation and coding scheme\\
		MCL&	Maximum coupling loss\\
		MIL&	Maximum isotropic loss\\
		MPL&	Maximum path loss\\
		Msg2&	Message 2 for random access response over PDSCH \\
		Msg3&	Message 3 for scheduled UL transmission over PUSCH \\
		Msg4&	Message 4 for contention resolution PDCCH or PDSCH\\
		NB-IoT& Narrowband Internet-of-Things\\
		NR&	New Radio\\
		OFDM &	Orthogonal frequency-division multiplexing \\
		PDCCH&	Physical downlink control channel  \\
		PDSCH&	Physical downlink shared channel \\
		PRACH&	Physical random-access channel \\
		PRB&	Physical resource block \\
		PUCCH&	Physical uplink control channel \\
		PUSCH&	Physical uplink shared channel \\
		Rx&	Receiver \\
		SCS&	Subcarrier spacing \\
		SSB&	Synchronization signal block \\
		TBS&	Transport block size  \\
		TDD&	Time division duplex \\
		TDRA&	Time domain resource allocation \\
		Tx&	Transmitter  \\
		UE&	User equipment  \\
		UL&	Uplink \\
			\bottomrule
	\end{tabular}}
	
	\label{Abbr}
\end{table}

\section{REDUCED CAPABILITY NEW RADIO DEVICES (NR-REDCAP)}
The use cases envisioned for RedCap include industrial wireless sensor network (IWSN), video surveillance cameras, and wearables (e.g., smart watches, rings, eHealth related devices, medical monitoring devices, etc.).  The specific requirements of these use cases are summarized in Table \ref{Requrements}.  As can be seen from Table \ref{Requrements}, the requirements on data rate, latency and reliability are diverse for RedCap use cases. Furthermore, these requirements differ significantly from the requirements for LPWAN use cases, currently addressed by LTE-M and NB-IoT. Thus, NR-RedCap is not intended for LPWAN use cases and is mainly intended ``mid-range" IoT market segment.

In addition to the use case specific requirements in Table~\ref{Requrements}, the following generic requirements are common to all RedCap use cases \cite{SID_redcap}:
\begin{itemize}

\item 	Lower device cost and complexity as compared to high-end eMBB and URLLC devices of Release-15/Release-16.
\item 	Smaller device size or compact form factor, and
\item 	Support deployment in all FR1/FR2 bands for FDD and TDD.
\end{itemize}

In order to meet the above generic requirements, and more specifically the one on device complexity and device size, the following features have been considered in the RedCap study item \cite{SID_redcap}: 

\begin{enumerate}[label=\alph*)]

\item	Reduced number of UE receiver (Rx) and/or transmitter (Tx) branches, 
\item	UE bandwidth reduction,
\item Half-duplex FDD,
\item Relaxed UE processing time,
\item 	Relaxed UE processing capability.
\end{enumerate}

The complexity reduction features which are expected to have the largest impact on coverage performance are (a) reduced number of UE Rx/Tx branches and (b) UE bandwidth reduction. Therefore, in what follows, we describe these features in more detail. More details on features (c), (d) and (e) are provided in TR 38.875 \cite{TR38875}.

The reduction of minimum number of Rx and/or Tx branches relative to that of a reference Rel-15 NR UE will lower the cost and complexity of the RedCap UEs. The reference NR UE supports 2Rx/1Tx branches in FR1 FDD bands, 4Rx/1Tx branches  in FR1 TDD bands, and 2Rx/1Tx branches in FR2 bands \cite{TR38875}. For RedCap UEs, the configuration for Rx and Tx branches that were considered are 1Rx/1Tx and 2Rx/1Tx, in both FR1 and FR2. Furthermore, carrier aggregation is not considered. The cost reduction, relative to that of the reference NR UE and in terms of modem bill of materials, from reducing the minimum number of Rx branches is summarized in Table \ref{CostRx} \cite{TR38875}. In FR1, the reduction of number of Rx branches is also beneficial in terms of reducing the device size. In FR2, however, the reduction of number of Rx branches may not provide much benefit in terms of reducing the device size as the antenna separation is in the order of the wavelength.

\begin{table}[!t]
	\centering 
	\caption{\small Use case specific requirements for RedCap.}
	\resizebox{\columnwidth}{!}{ 
		\begin{tabular}{ |c|c|c|c|c| }  
			\hline 
			&\multirow{2}{5em}{ IWSN (non-safety)}&	\multirow{2}{5em}{IWSN (safety)}&	\multirow{2}{5em}{Video surveillance}&	Wearables \\
			&&&&\\
			\hline
			\multirow{2}{7.5em}{Data rate (reference bit rate)}&	UL: < 2 Mbps& 	UL: < 2 Mbps&	UL$^{1}$: 2-4 Mbps&	\multirow{2}{6em}{UL$^{2}$: 2-5 Mbps,	DL$^{3}$: 5-50 Mbps} 
			\\ &&&&\\ &&&&\\	\hline
			Latency	& < 100 ms &	5-10 ms	& < 500 ms&	- \\	\hline
			Battery life&	Few years&	-&	-&	1-2 weeks\\	\hline
			Reliability&	99.99\% &	99.99\%	&99\%-99.9\% &	- \\	\hline
			
			\multicolumn{5}{|l|} {Note 1: High-end video e.g. for farming would require 7.5-25 Mbps.}\\
			\multicolumn{5}{|l|} {Note 2: Peak data rate of the wearables is up to 50 Mbps for uplink.}\\
			\multicolumn{5}{|l|} {Note 3: Peak data rate of the wearables is up to 150 Mbps for downlink.}
			\\	\hline
	\end{tabular}}
	
	\label{Requrements}
	
\end{table}

\begin{table}[!t]
	
	\centering
	\caption{\small Estimated relative UE cost reduction for reduced number of UE Rx branches.}
	\resizebox{\columnwidth}{!}{
		\begin{tabular}{ |c|c|c|c|c| } 
			\hline
			
			\multirow{2}{7em}{Reduced number of Rx branches}& \multirow{2}{5.5em}	{FR1 FDD
				(2Rx to 1Rx)} &	\multirow{2}{5.5em}{FR1 TDD
				(4Rx to 2Rx)}&	\multirow{2}{5.5em}{FR1 TDD 
				(4Rx to 1Rx)}&	\multirow{2}{5.5em}{FR2 (2Rx to 1Rx)}
			
			\\ &&&&\\	\hline
			Case 1	& ~26\%&	~31\%	&~46\%&	~31\% \\	\hline
			Case 2 &	~37\%	& ~40\% &	~60\%	&~40\%\\	\hline
			\multicolumn{5}{|l|} {Case 1: 	Total cost reduction (without DL MIMO layer reduction).}\\
			\multicolumn{5}{|l|} {	\multirow{2}{30em} {Case 2:	Total cost reduction (with DL MIMO layer reduction, i.e. number of MIMO layer equal to number of Rx branches). }}	\\
			\multicolumn{5}{|l|} {}
			\\\hline
	\end{tabular}}
	
	\label{CostRx}
\end{table}

\begin{table}[!t]
	\centering
	\caption{\small Estimated relative UE cost reduction for reduced maximum UE bandwidth.}
	\resizebox{\columnwidth}{!}{
		\begin{tabular}{ |c|c|c|c|c| } 
			\hline
			
			\multirow{2}{5.5em}{Reduced UE bandwidth}& \multirow{2}{6em} {FR1 FDD
				(100 MHz to 20 MHz)} &	\multirow{2}{6em}{FR1 TDD
				(100 MHz to 20 MHz)}&	\multirow{2}{6em}{FR2
				(200 MHz to 100 MHz)}&	\multirow{2}{6em}{FR2
				(200 MHz to 50 MHz)}	\\ &&&&\\ &&&&\\	\hline
			Total cost reduction &	~32\%	 &~33\%	&~16\%	&~24\%
			\\\hline
	\end{tabular}}
	
	\label{CostBW}
\end{table}

In addition to the reduction in number of Rx branches, UE bandwidth reduction is another important feature that can considerably bring down the cost and complexity of the RedCap UE. For the estimation of relative cost/complexity saving due to UE bandwidth reduction, a Release 15 NR UE is used as a reference. The maximum bandwidth capability of the reference UE is assumed to be 100 MHz in FR1 and 200 MHz in FR2, for both uplink and downlink. For RedCap UEs, the bandwidth reduction options considered during the study item \cite{SID_redcap} are 20 MHz in FR1 and 50 or 100 MHz in FR2. The cost reduction, relative to that of the reference NR UE, is summarized in Table \ref{CostBW} \cite{TR38875}.

As shown in Table \ref{CostRx} and Table \ref{CostBW} , the reduction of number of Rx branches and UE bandwidth will lead to cost saving benefits for the RedCap UEs. The drawback, however, is that the performance and consequently the coverage of the UEs can be negatively impacted. In the following sections, we evaluate the coverage impacts that entail from the use of these complexity reduction features. In addition to the complexity reduction features, the coverage analysis in FR1 also takes into consideration reduced antenna efficiency due to size limitations for devices such as wearables. The antenna efficiency loss is limited to 3 dB, and is considered for both uplink and downlink channels in the link budget evaluations.

\section{LINK LEVEL SIMULATIONS}

In order to evaluate the impact of the UE complexity reduction  on coverage of RedCap physical channels, as the first step we have performed link-level simulations (LLS) to obtain the required SINR for the physical channels under performance target for the both reference UEs and RedCap UEs. Then, the outcomes of the LLSs are used to perform the link budget evaluation to find coverage limiting channels.
 As it is expected that the coverage of a physical channel is affected by complexity reduction differently in different frequency bands, we have performed the LLSs for three different scenarios: 

\begin{enumerate}
	\item FR1, Rural with the carrier frequency of 0.7 GHz,
	\item FR1, Urban with the carrier frequency of 2.6 GHz,
	\item FR2, Indoor with the carrier frequency of 28 GHz.
\end{enumerate}

Any of the UL and DL initial access messages or physical channels can be potentially coverage limiting for RedCap UEs, therefore, we have considered LLSs for the following messages and channels \cite{dahlman20185g}:
\begin{itemize}
\item	Synchronized signal block (SSB), including primary SS (PSS), secondary SS (SSS) and physical broadcast channel (PBCH), is periodically transmitted on DL to initial cell search (in this paper mainly consider PBCH), and carries the information that UE needs in order to connect to the network,
\item	Physical random-access channel (PRACH), is used by UE for transmission of preamble over UL, 
\item	Message 2 (Msg2) or random-access response, is transmitted on DL for indicating reception of the preamble and sending time alignment information, 
\item	Message 3 (Msg3) is used by UE to transmit information such as a device identity that is needed for the next message over PUSCH,
\item	Message 4 (Msg4) transfers the UE to the connected state,
\item	Physical downlink control channel (PDCCH), is mainly used for transmission of control information such as scheduling decisions,  
\item	Physical downlink shared channel (PDSCH), is mainly used as the main transmission of DL unicast data,
\item	Physical uplink control channel (PUCCH) is used by UE to send information such as acknowledgments and channel-state reports,  
\item	Physical downlink shared channel (PUSCH), is the uplink counterpart of PDSCH. 

\end{itemize}

Our general simulation assumptions for the reference UE are listed in Table \ref{LS_assum}. To investigate the impact of the complexity reduction on the BLER performance of RedCap UEs, we have also performed the LLSs for RedCap UEs considering the parameters shown in Table \ref{LS_assum}, except that the UE bandwidth and the number of Rx branches are reduced as reported in Table \ref{LS_assum_redcap}.

Our channel-specific assumptions, the required performance targets such as the BLER performance are discussed separately for each channel in the following sections. Moreover, the SINR requirements for meeting BLER targets for different channels and signals in FR1 and FR2 are summarized in Tables \ref{SNR_700}-\ref{SNR_28}.    

\begin{table}[!t]
	\centering
	\caption{\small Link-level simulations assumptions for reference UE.}
	\resizebox{0.9\columnwidth}{!}{
		\begin{tabular}{@{}ll@{}}
			\toprule
			
			Carrier frequencies	& Rural: 700 MHz (FDD) \\
			&	Urban: 2.6 GHz (TDD) \\
			&Indoor: 28 GHz (TDD)\\ \hline
			BWP BW &	Rural: 20 MHz\\
			&	Urban: 100 MHz\\
			&	Indoor: 100 MHz \\ \hline
			SCS &	Rural: 15 kHz\\
			&	Urban: 30 kHz\\
			&	Indoor: 120 kHz \\ \hline
			Frame structure for TDD &	Urban: DDDDDDDSUU (S: 6D:4G:4U)\\
			&	Indoor: DDDSU (S: 10D:2G:2U) \\ \hline
			Number of gNB TX chains &	Urban: 4\\
			&	Rural, Indoor: 2 \\ \hline
			Number of gNB RX chains & 	Urban: 4 \\
			&	Rural, Indoor: 2 \\ \hline
			Number of UE TX chains &	Rural, Urban, Indoor: 1 \\ \hline
			Number of UE RX chains &	Urban: 4 \\
			&Rural, Indoor: 2 \\ \hline
			Channel model	& Rural, Urban,: TDL-C, NLOS\\
			&	Indoor: TDL-A, NLOS \\ \hline
			UE antenna correlation &	Rural, Urban, Indoor: Low \\ \hline
			Delay spread &	Rural, Urban: 300 ns \\
			&	Indoor: 30 ns \\ \hline
			UE velocity	 &Rural, Urban, Indoor: 3 km/h \\ 
			
			\bottomrule
	\end{tabular}}
	
	\label{LS_assum}
\end{table}

\begin{table}[!t] 
	\centering 
	\begin{small}
		\caption{\small Link-level simulations assumptions for RedCap.}
		\resizebox{0.85\columnwidth}{!}{
			\begin{tabular}{@{}ll@{}}
				\toprule
				\footnotesize	BWP BW &	\footnotesize Rural: 20 MHz \\
				\footnotesize	&	\footnotesize Urban: 20 MHz\\
				\footnotesize 	&	\footnotesize Indoor: 100 MHz
				\\ 
				\footnotesize 	Number of UE RX chains & \footnotesize	Rural, Urban, Indoor: 2 and 1\\

				\bottomrule
		\end{tabular}}
		
		\label{LS_assum_redcap}
		
	\end{small}
\end{table}

\subsection{SSB}
Based on the assumptions reported in Table 7, we have performed the LLSs for both reference UE-SSB and RedCap-SSB.

\begin{table}[!t]
	\centering 
	\caption{\small Channel-specific parameters for SSB.}
	\resizebox{0.8\columnwidth}{!}{
		\begin{tabular}{@{}ll@{}} 
			
			\toprule 
			\footnotesize	Channel & \footnotesize	Assumptions \\ \hline
			\footnotesize	SSB	  & \footnotesize (Residual) frequency offset (UE): 0.1 ppm \\
			\footnotesize & \footnotesize	SS burst set periodicity: 20 ms\\
			\footnotesize &\footnotesize	Precoder: Precoder cycling\\
			\footnotesize &\footnotesize	Number of transmissions (shots): 4\\
			\footnotesize &\footnotesize	BLER target: 1\% \\
			
			\bottomrule
	\end{tabular}}
	
	\label{SSB_LS}
\end{table}

\begin{figure}[!t]
	\centering
	\includegraphics[width=\columnwidth]{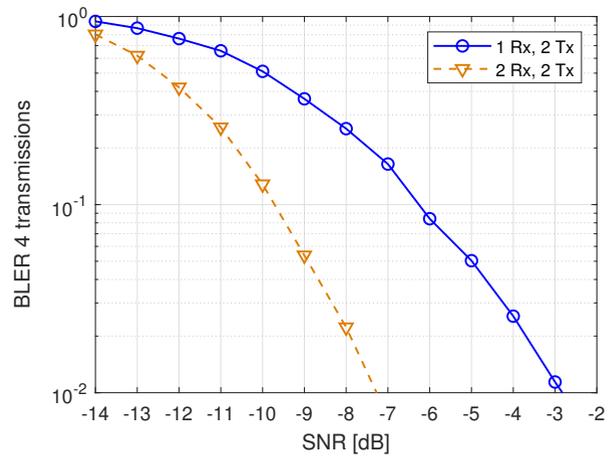}
	\caption{BLER performance of SSB, 700 MHz.} 
	\label{SSB700}
\end{figure}

\begin{figure}[!t]
	\centering
	\includegraphics[width=\columnwidth]{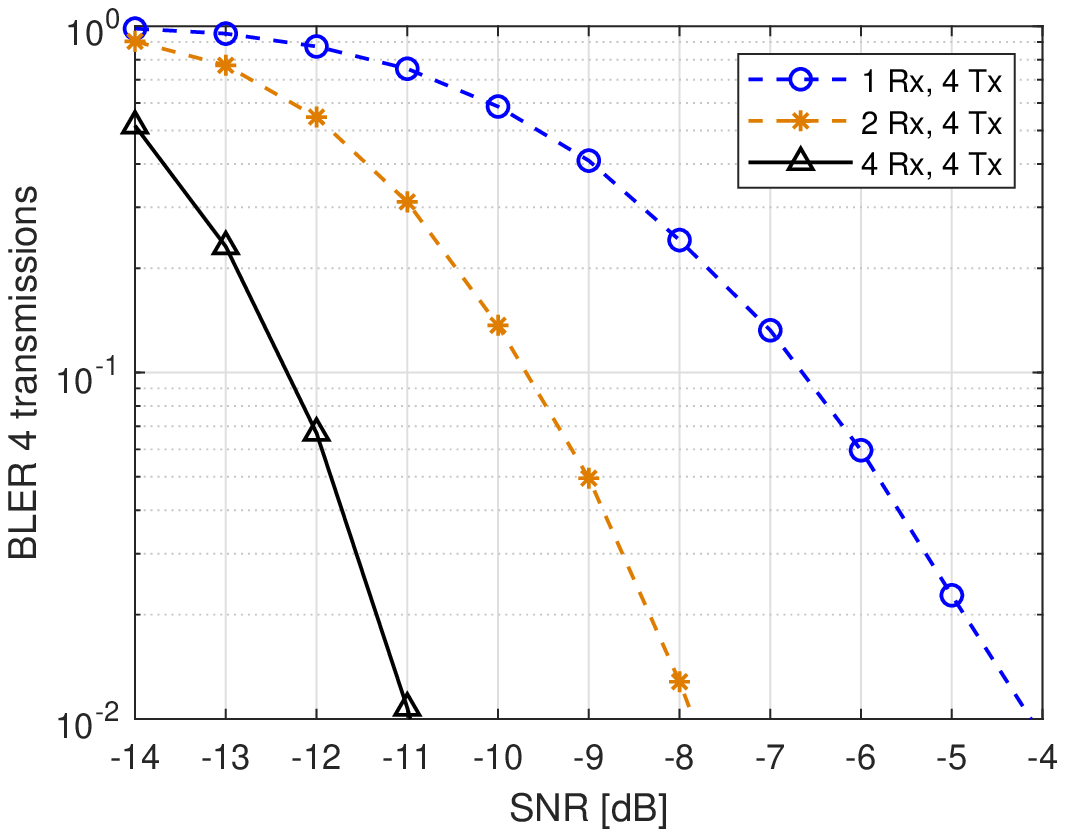}
	\caption{BLER performance of SSB, 2.6 GHz.} 
	\label{SSB2.6}
\end{figure}

\begin{figure}[!t]
	\centering
	\includegraphics[width=\columnwidth]{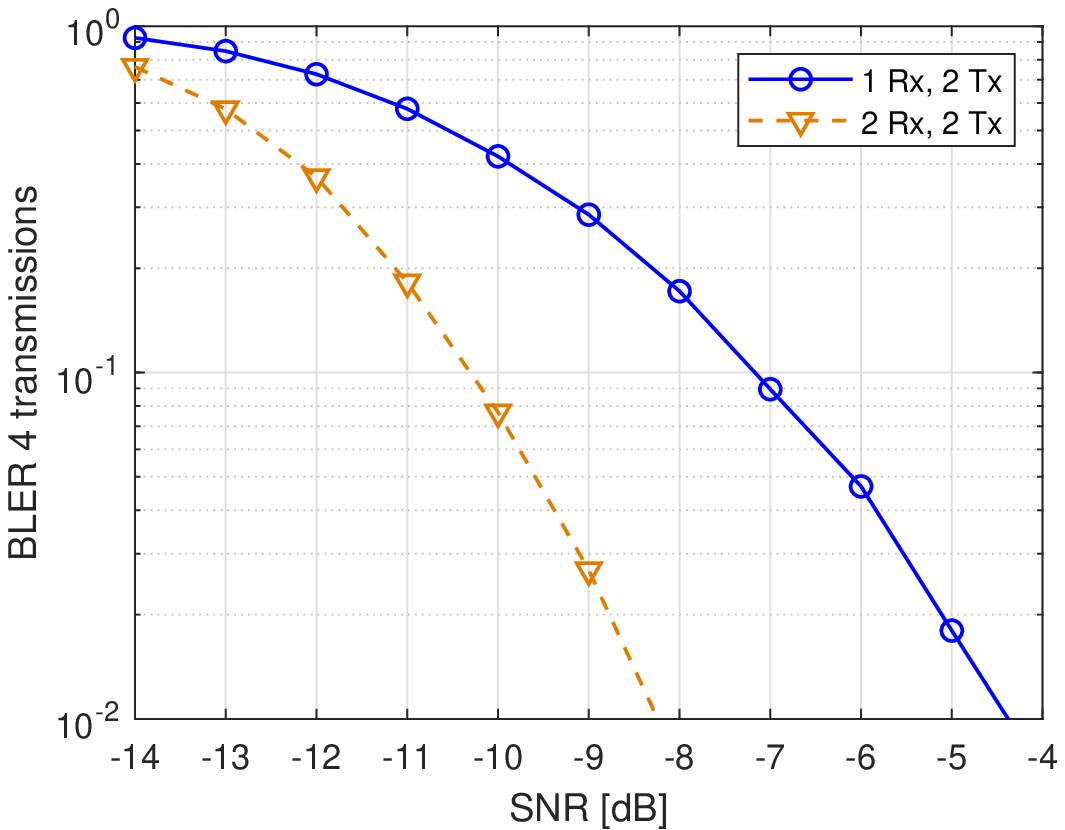}
	\caption{BLER performance of SSB, 28 GHz.} 
	\label{SSB28}
\end{figure}

Our simulation results are shown in Figures \ref{SSB700}-\ref{SSB28} for Rural, Urban, and Indoor scenarios, respectively. For Rural scenario at carrier frequency of 700 MHz, the performance (at 1\% BLER) degrades by 4.4 dB considering the complexity reductions for RedCap UEs.

Based on the results shown in Figure \ref{SSB2.6}, the performance losses for PBCH (after 4 transmissions, at 1\% BLER) incurred from reducing the number of receiver branches for a RedCap UE with respect to the reference NR UE are 3.0 dB and 6.9 dB for a 2 Rx and 1 Rx RedCap UE, respectively, for Urban scenario. For the Indoor scenario in the FR2 band, as it is shown in Figure \ref{SSB28}, reducing the Rx branches to 1 the BLER performance degrades by 3.7 dB at 1\% BLER.

\subsection{PRACH}

\begin{table}[!t]
	
	\caption{\small Channel-specific parameters for PRACH.}
	\resizebox{\columnwidth}{!}{
		\begin{tabular}{@{}ll@{}}
			\toprule
			Channel &	Assumptions \\ \hline
			PRACH &	PRACH format: \\
			& \,\,\,	- Rural: Format 0 (1.25 KHz SCS), BW = 1.04875 MHz \\
			&	\,\,\, - Urban: Format B4 (15 KHz SCS), BW = 2.085 MHz\\
			&\,\,\,	- Indoor: Format B4 (120 KHz SCS), BW = 16.68 MHz \\ 
			
			&	Number of transmissions: 1 \\
			&	Rx combining: non-coherent combining of branches \\
			&	Propagation delay (RTT):\\
			&\,\,\,	- Rural: uniformly distributed [0, 23] µs (ISD 6 km)\\
			&\,\,\,	- Urban: uniformly distributed [0, 2.7] µs (ISD 700 m)\\
			&\,\,\,	- Indoor: uniformly distributed [0, 0.077] µs (ISD 20 m)\\
			
			&	Frequency error: 0.1 ppm at UE, none at gNB\\
			&	Performance target: 10\% and 1\% missed detection \\
			& \hspace{2.9cm}at 0.1\% false alarm probability \\

			\bottomrule
	\end{tabular}}
	
	\label{PRACH_LS}
\end{table}

\begin{figure}[!t]
	\centering
	\includegraphics[width=\columnwidth]{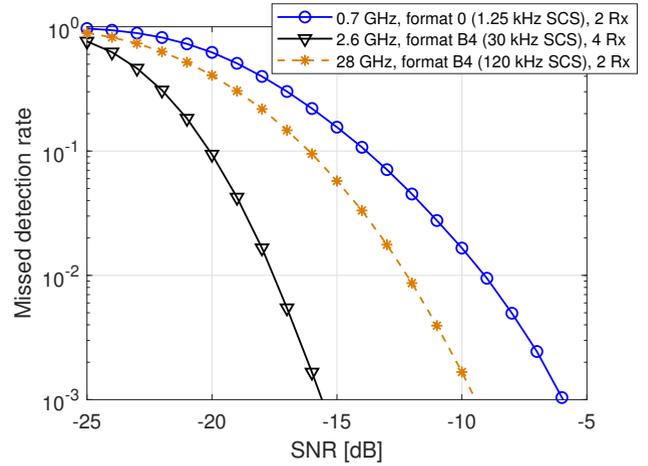}
	\caption{Missed detection rate of PRACH.} 
	\label{PRACHMD}
\end{figure}

Table \ref{PRACH_LS} represents our assumptions for LLS of PRACH. The miss detection rate for the PRACH of Rural, Urban, and Indoor scenarios are shown in Figure \ref{PRACHMD}.  In the uplink, the number of Tx branches is the same at the reference NR UE and the RedCap UE. Furthermore, as shown in Table 8, the PRACH BW for each PRACH occasion in the frequency domain is less than that of the RedCap UE BW in all the considered scenarios. Therefore, the link performance of RedCap-PRACH is identical to that of the reference UE-PRACH.

\subsection{Msg2}

Our simulation assumptions for performing Msg2 LLSs, are shown in Table \ref{Msg2_LS}. For Msg2 we have considered the payload size of 9 bytes and MCS index of 0 from Table 5.1.3.1-1 in \cite{TS38214}. We have also considered TBS scaling of 0.25, so that a smaller TBS can be assigned to a given MCS and a given number of PRBs, by considering a TBS scaling factor in computing $N_\text{info}$ as \cite{TS38214}:

\begin{equation}
N_\text{info}=SN_{RE}Q_mv,
\end{equation}
where $S$, $N_{RE}$, $R$, $Q_m$, and $v$ are the scaling factor, the number of available resource elements, code rate, modulation order, and the number of transmission layers, respectively.

\begin{table}[!t]
	
	\caption{\small Channel-specific parameters for Msg2.}
	\resizebox{\columnwidth}{!}{
		\begin{tabular}{@{}ll@{}}
			\toprule
	Channel &	Assumptions \\ \hline
		Msg2 & 	FDRA: 12 PRBs (considering TBS scaling factor is 0.25)\\
	&	TDRA: 12 OFDM symbols \\
	&	Waveform: CP-OFDM \\
	&	DMRS: Type I, 3 DMRS symbol, no multiplexing with data \\
	&	Payload/MCS index: 9 bytes/MCS0  \\
	&	Number of transmissions: No HARQ \\
	&	Rx combining: MRC\\
	&	Precoder: Precoder cycling, PRB bundle size of 2 \\
	&	BLER target: 10\% \\
	
			\bottomrule
	\end{tabular}}
	\label{Msg2_LS}
\end{table}

\begin{figure}[!t]
	\centering
	\includegraphics[width=\columnwidth]{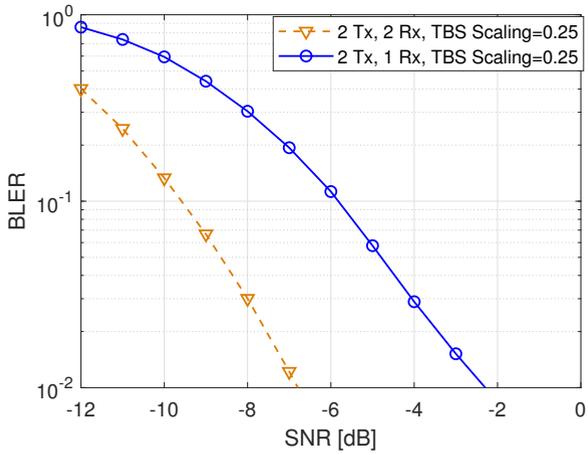}
	\caption{BLER performance of Msg2, 700 MHz.} 
	\label{Msg2_700}
\end{figure}

\begin{figure}[!t]
	\centering
	\includegraphics[width=\columnwidth]{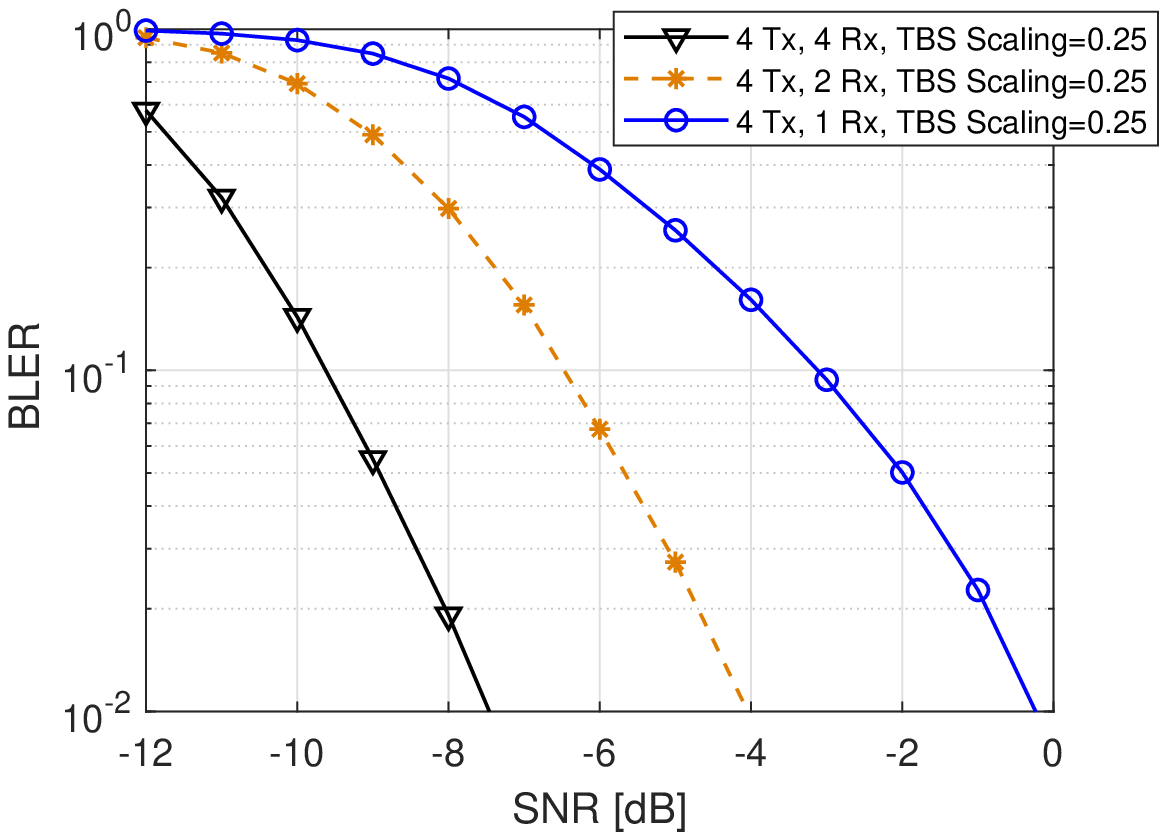}
	\caption{BLER performance of Msg2, 2.6 GHz.} 
	\label{Msg2_2.6}
\end{figure}

\begin{figure}[!t]
	\centering
	\includegraphics[width=\columnwidth]{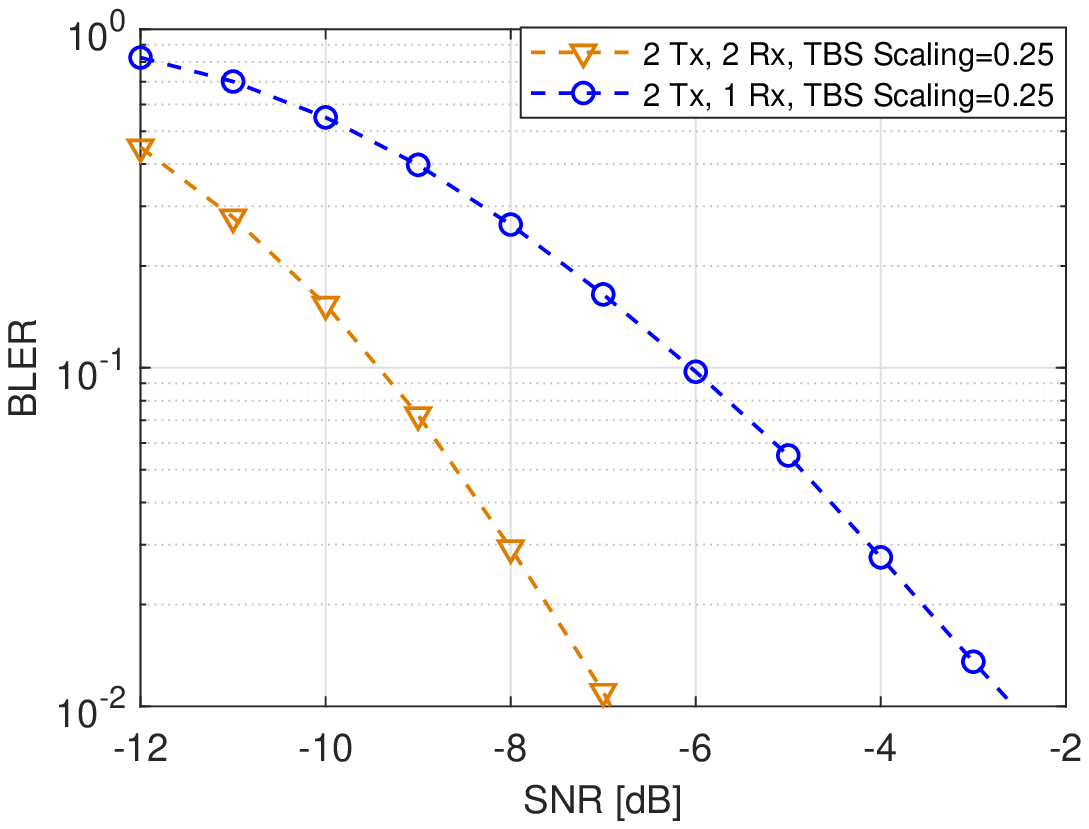}
	\caption{BLER performance of Msg2, 28 GHz.} 
	\label{Msg2_28}
\end{figure}

Figures \ref{Msg2_700}-\ref{Msg2_28} show the BLER performance of Msg2 at carrier frequencies of 700 MHz, 2.6 GHz and 28 GHz, respectively. As it is shown in Figure \ref{Msg2_700}, at BLER performance of 10\%, by reducing the number of UE Rx branches to 1, Msg2 performance is degraded by 6.5 dB for Rural case.

Based on our results shown in Figure \ref{Msg2_2.6}, at carrier frequency of 2.6 GHz and BLER performance of 10\%, Msg2 performance is respectively degraded by 3.1 dB and 3.4 dB for reducing the number of UE Rx branches from 4 to 2 and from 2 to 1. As it is shown in  Figure \ref{Msg2_28}, at carrier frequencies of 28 GHz and BLER performance of 10\%, reducing the number of UE Rx branches to 1, Msg2 performance is degraded by 3.8 dB.

\subsection{Msg3}

\begin{table}[!t]
	
	\caption{\small Channel-specific parameters for Msg3.}
	\resizebox{\columnwidth}{!}{
		\begin{tabular}{@{}ll@{}}
			\toprule
			Channel &	Assumptions \\ \hline
		Msg3 &	FDRA: 2 PRBs \\ 
		& TDRA: 14 OFDM symbols \\ 
		& Waveform: DFT-s-OFDM \\ 
	&	DMRS: Type I, 3 DMRS symbol, no multiplexing with data \\ 
	&	Payload/MCS index: 56 bits/MCS0  \\ 
	&	Number of transmissions: No HARQ \\ 
	&	Rx combining: MRC \\ 
	&	No frequency hopping \\ 
	&	BLER target: 10\% \\ 
		
			\bottomrule
	\end{tabular}}
	\label{Msg3_LS}
\end{table}

\begin{figure}[!t]
	\centering
	\includegraphics[width=\columnwidth]{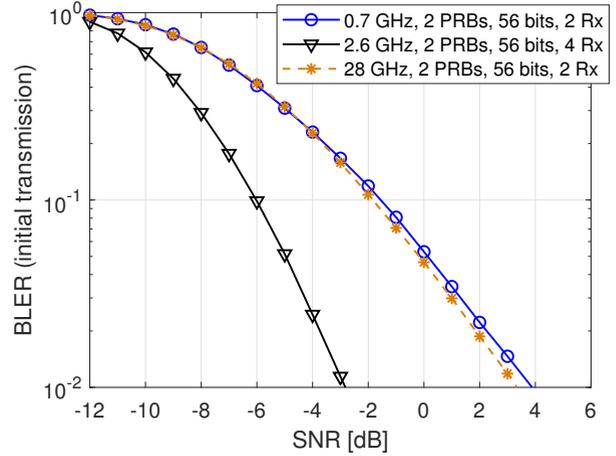}
	\caption{BLER performance of Msg3.} 
	\label{Msg3}
\end{figure}

Table \ref{Msg3_LS} shows the assumptions for LLS of Msg3. The BLER performance of Msg3 is shown in Figure \ref{Msg3}. Similar to PRACH, the Msg3 performance of the RedCap UE is the same as that of the reference NR UE. This is because the BW of Msg3 is assumed to be smaller than the RedCap UE BW, and the number of the Tx branches of the RedCap UE is identical to that of the reference UE. 

\subsection{Msg4}

\begin{table}[!t]
	
	\caption{\small Channel-specific parameters for Msg4.}
	\resizebox{\columnwidth}{!}{
		\begin{tabular}{@{}ll@{}}
			\toprule
			Channel &	Assumptions \\ \hline
	Msg4	& FDRA (reference UE):\\
	& \,\,\,- Rural: 36 PRBs\\
   &  \,\,\,- Urban: 36 PRBs\\
	& \,\, - Indoor: 37 PRBs\\
	
&	TDRA: 12 OFDM symbols \\
&	Waveform: CP-OFDM \\
&	DMRS: Type I, 3 DMRS symbol, no multiplexing with data \\
&	Payload/MCS index: 130 bytes /MCS0  \\
	
&	Number of transmissions: No HARQ \\
&	Rx combining: MRC \\
&	Precoder: Precoder cycling; PRB bundle size of 2 \\
&	BLER target: 10\% \\		
			\bottomrule
	\end{tabular}}
	\label{Msg4_LS}
\end{table}

\begin{figure}[!t]
	\centering
	\includegraphics[width=\columnwidth]{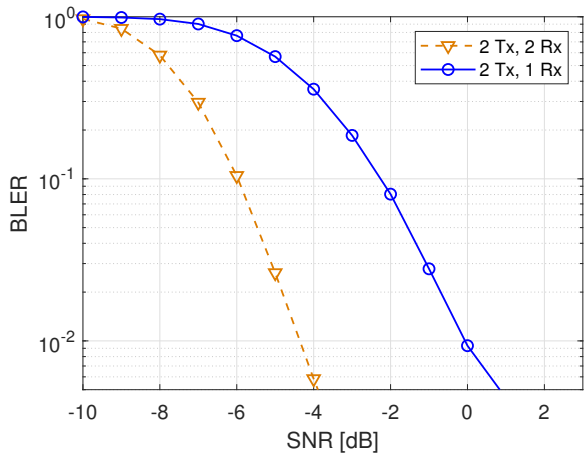}
	\caption{BLER performance of Msg4, 700 MHz.} 
	\label{Msg4_700}
\end{figure}

\begin{figure}[!t]
	\centering
	\includegraphics[width=\columnwidth]{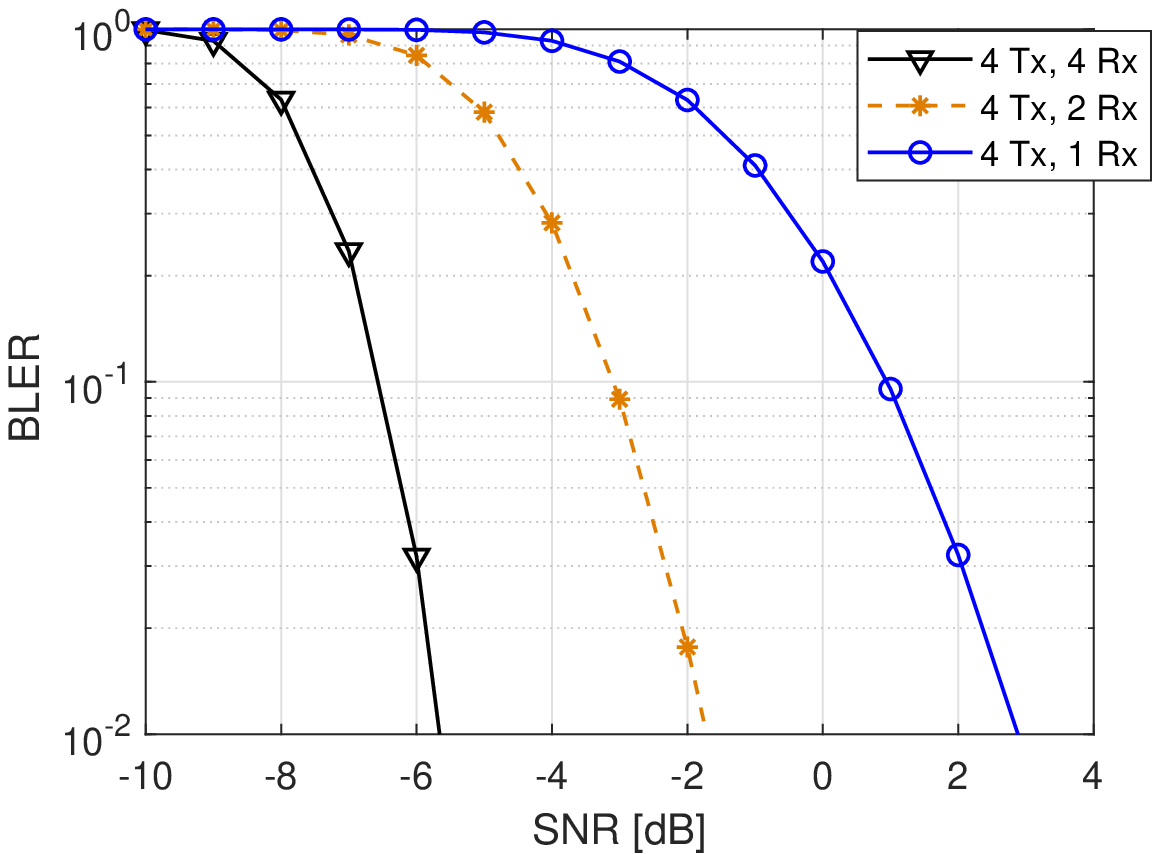}
	\caption{BLER performance of Msg4, 2.6 GHz.} 
	\label{Msg4_2.6}
\end{figure}

\begin{figure}[!t]
	\centering
	\includegraphics[width=\columnwidth]{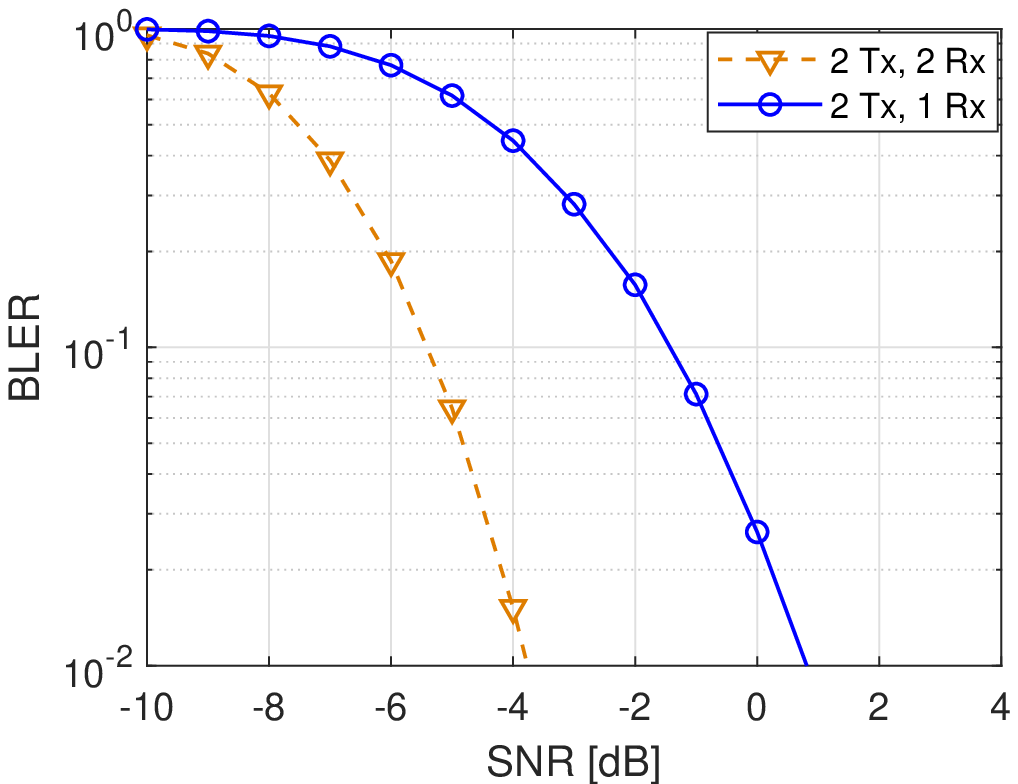}
	\caption{BLER performance of Msg4, 28 GHz.} 
	\label{Msg4_28}
\end{figure}

Our simulation assumptions for LLS of Msg4 are shown in 
Table \ref{Msg4_LS}. Figures \ref{Msg4_700}-\ref{Msg4_28} show the BLER performance of Msg4 at carrier frequencies of 700 MHz, 2.6 GHz and 28 GHz, respectively. Based on our simulation results in Figure \ref{Msg4_700}, by reducing the BW and the number of UE Rx branches to 1, Msg4 performance is degraded by 4.1 dB at carrier frequencies 700 MHz. As it is shown in Figure \ref{Msg4_2.6}, at carrier frequency of 2.6 GHz and BLER performance of 10\%, Msg4 performance is respectively degraded by 3.5 dB and 4 dB for reducing the number of Rx branches from 4 to 2 and from 2 to 1. For Indoor scenario, by reducing the BW and the number of UE Rx branches to 1, at BLER performance of 10\%, Msg4 performance is degraded by 4 dB.

\subsection{PDCCH}

\begin{figure}[!t]
	\centering
	\includegraphics[width=\columnwidth]{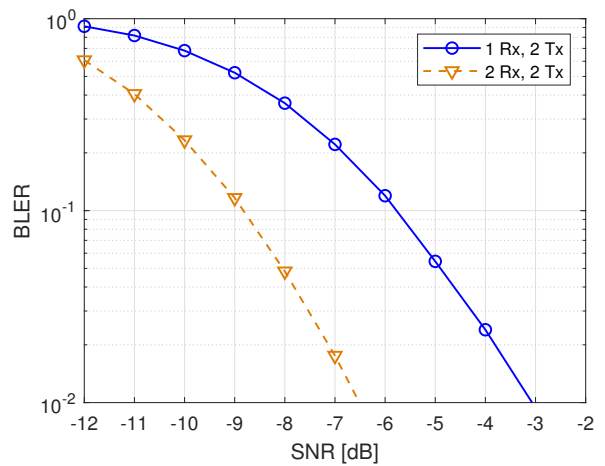}
	\caption{BLER performance of PDCCH, 700 MHz.} 
	\label{PDCCH_700}
\end{figure}

\begin{figure}[!t]
	\centering
	\includegraphics[width=\columnwidth]{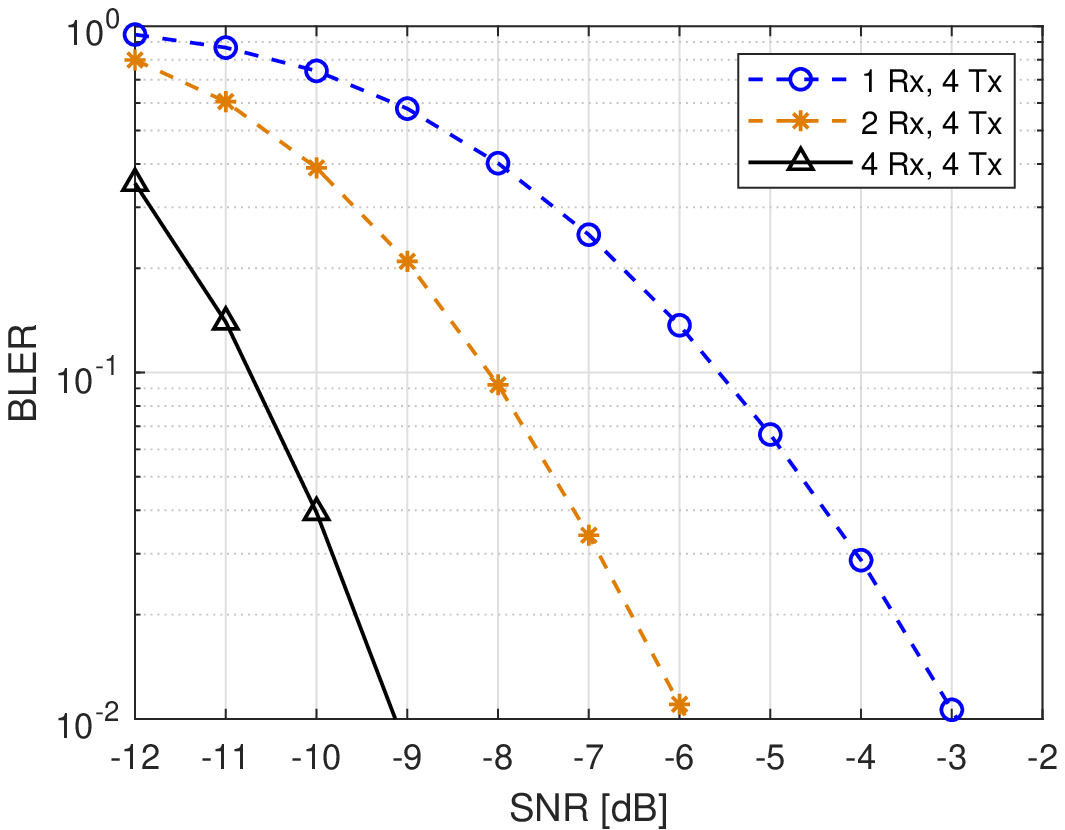}
	\caption{BLER performance of PDCCH, 2.6 GHz.} 
	\label{PDCCH_2.6}
\end{figure}

\begin{figure}[!t]
	\centering
	\includegraphics[width=\columnwidth]{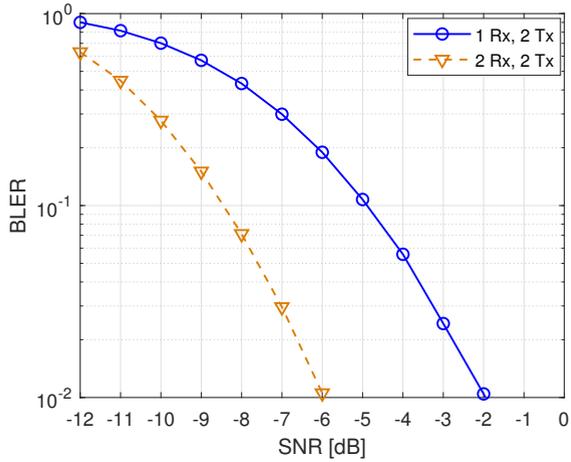}
	\caption{BLER performance of PDCCH, 28 GHz.} 
	\label{PDCCH_28}
\end{figure}

We have performed the LLS for PDCCH channel based on the assumptions reported in Table \ref{PDCCH_LS}, and our simulation results are shown in Figures \ref{PDCCH_700}-\ref{PDCCH_28}, respectively for carrier frequencies 700 MHz, 2.6 GHz, and 28 GHz.

\begin{table}[!t]
	
	\caption{\small Channel-specific parameters for PDCCH.}
	\resizebox{\columnwidth}{!}{
		\begin{tabular}{@{}ll@{}}
			\toprule
			Channel &	Assumptions \\ \hline
		PDCCH &	DCI payload size: 40 bits+CRC \\
		& Aggregation level (AL): 16 \\
		& CORESET: 2 symbols and 48 PRBs \\
		& Precoding: Precoder cycling at CCE level (REG bundle = 6) \\
		& BLER target: 1\% \\
		
			\bottomrule
	\end{tabular}}
	
	\label{PDCCH_LS}
\end{table}

For Rural scenario at carrier frequency of 700 MHz, the performance (at \%1 BLER) degrades by 3.5 dB considering the complexity reductions for RedCap UEs.

The performance losses for PDCCH (at 1\% BLER) incurred from reducing the number of receiver branches for a RedCap UE with respect to the reference NR UE are 3.2 dB and 6.2 dB for a 2 Rx and 1 Rx RedCap UE, respectively at carrier frequency of 2,6 GHz. In FR2 band at carrier frequency of 28 GHz, the performance loss is 3.9 dB by reducing the number of Rx branches to 1 for RedCap UE.

\subsection{PDSCH}

Table \ref{PDSCH_LS} show our assumptions for LLS of PDSCH. It is worth to mention that our assumptions on data rate target is based on agreements from \cite{SID_redcap} and for the given number of PRBs, we have selected the smallest MCS index, from table 5.1.3.1-1 in \cite{TS38214}, that satisfies our data rate constraints.

\begin{table}[!t]
	
	\caption{\small Channel-specific parameters for PDSCH.}
	\resizebox{\columnwidth}{!}{
		\begin{tabular}{@{}ll@{}}
			\toprule
			Channel &	Assumptions \\ \hline
		PDSCH &	FDRA (reference UE): \\
		& \,\,\,- Urban: 200 PRBs \\
	&	\,\,\,- Indoor: 60 PRBs \\
	&	\,\,\,- Rural: 40 PRBs\\
	&	FDRA (RedCap UE):\\
	&	\,\,\,- Urban: 51 PRBs\\
	& \,\,\,- Indoor: 30 PRBs\\
	&\,\,\,- Rural: 40 PRBs\\
		
	&	TDRA: 12 OFDM symbols\\
	&	Waveform: CP-OFDM\\
	&	DMRS: Type I, 2 DMRS symbol, no multiplexing with data\\
	&	Target data rate/TBS/MCS (reference UE): \\
	&	\,\,\,- Urban: 10 Mbps/TBS =5640/MCS0  \\
	&	\,\,\,- Indoor: 25 Mbps/TBS = 3624/MCS3\\
	&	\,\,\,- Rural: 1 Mbps/TBS = 1128/MCS0\\
	&	Target data rate/TBS/MCS (RedCap UE):\\
	&	\,\,\,- Urban: 10 Mbps/TBS = 1480/MCS0\\
	&	\,\,\,- Indoor: 25 Mbps/TBS =3240/MCS6\\
	&	\,\,\,- Rural: 1 Mbps/TBS = 1128/MCS0\\
		
	&	Number of transmissions: No HARQ\\
	&	Rx combining: MRC\\
	&	Precoder: Precoder cycling; PRB bundle size of 2\\
	&	BLER target: 10\% \\
					
			\bottomrule
	\end{tabular}}
	\label{PDSCH_LS}
\end{table}

\begin{figure}[!t]
	\centering
	\includegraphics[width=\columnwidth]{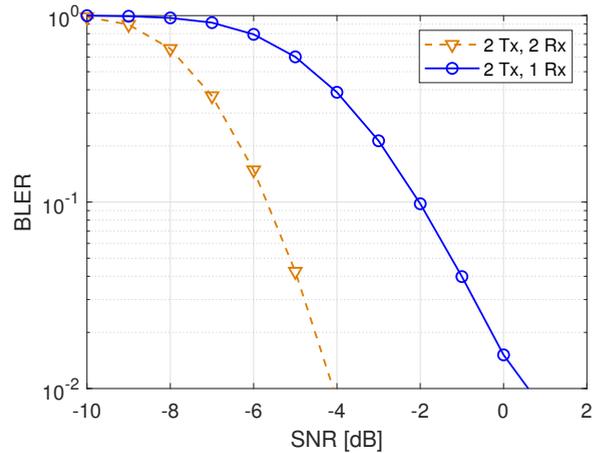}
	\caption{BLER performance of PDSCH, 700 MHz.} 
	\label{PDSCH_700}
\end{figure}

\begin{figure}[!t]
	\centering
	\includegraphics[width=\columnwidth]{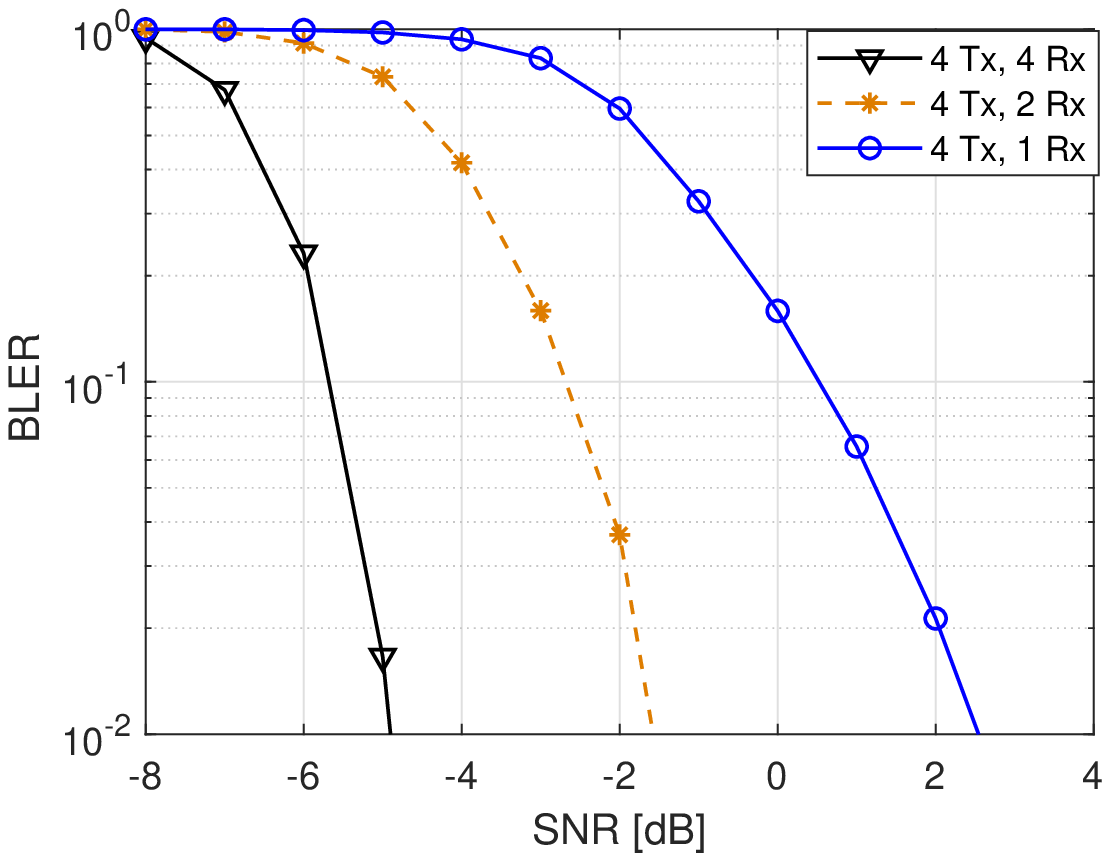}
	\caption{BLER performance of PDSCH, 2.6 GHz.} 
	\label{PDSCH_2.6}
\end{figure}

\begin{figure}[!t]
	\centering
	\includegraphics[width=\columnwidth]{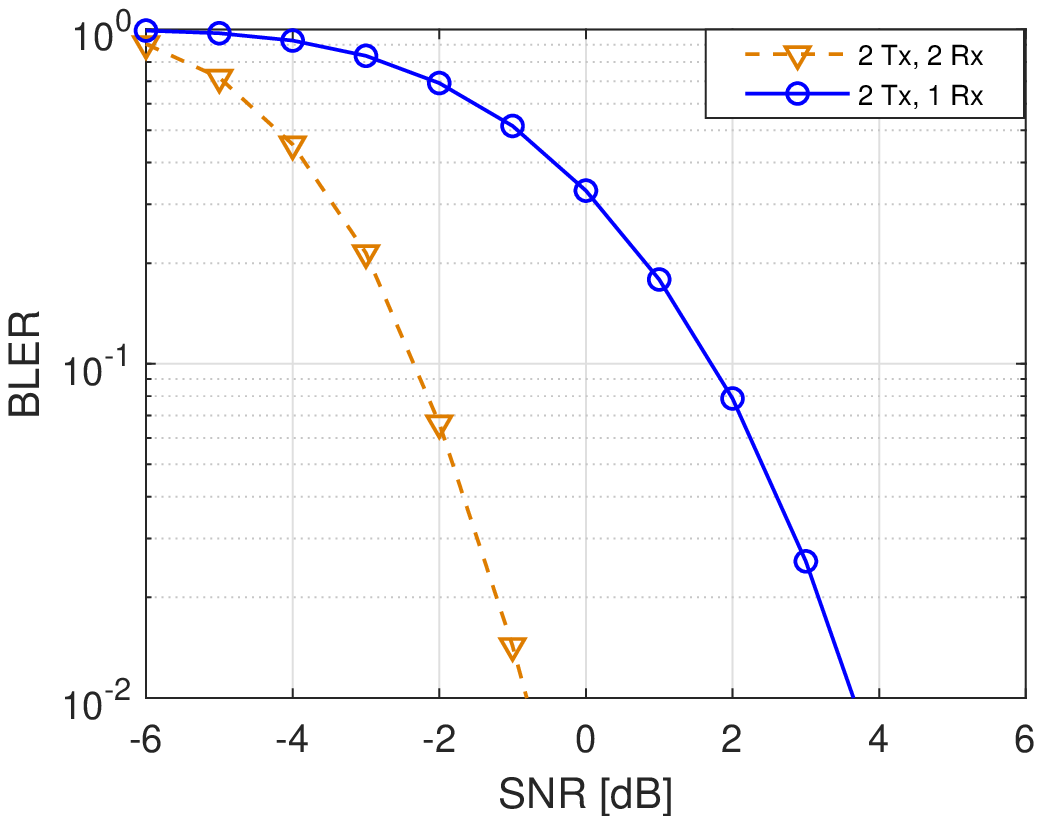}
	\caption{BLER performance of PDSCH, 28 GHz.} 
	\label{PDSCH_28}
\end{figure}

The BLER performances for PDSCH at carrier frequencies of 700 MHz, 2.6 GHz and 28 GHz are show in Figures \ref{PDSCH_700}-\ref{PDSCH_28}, respectively. As it is shown in these figures, at the carrier frequency of 700 MHz and 10\% BLER performance, by reducing the number of UE Rx branches to 1,  the performance of PDSCH is degraded by 3.8 dB.

As it is shown in Figure \ref{PDSCH_2.6}, at the carrier frequency of 2.6 GHz and BLER performance of 10\%, PDSCH performance is respectively degraded by 3 dB and 3.2 dB for reducing the number of Rx branches from 4 to 2 and from 2 to 1. As it is shown in Figure \ref{PDSCH_28}, For a RedCap UE with 1 Rx branch and operating at the carrier frequency of 28 GHz the PDSCH performance is 4 dB worse than that of the reference UE at 10\% BLER.

\subsection{PUCCH}

\begin{table}[!t]
	
	\caption{\small Channel-specific parameters for PUCCH.}
	\resizebox{\columnwidth}{!}{
		\begin{tabular}{@{}ll@{}}
			\toprule
			Channel &	Assumptions \\ \hline
		PUCCH	&FDRA: 1 PRB \\
		&TDRA: 14 OFDM symbols\\
		&Payload and format:\\
	&	\,\,\,- 2 bits (A/N) for format 1\\
	&	\,\,\,- 4/11/22 bits (A/N+SR/UCI) for format 3 \\
	&	Frequency hopping: At UL BWP edge\\
	&	DMRS:\\
	&	\,\,\,- Format 1: every even symbol according to the specification\\
	&	\,\,\,- Format 3: Additional DMRS configured (4 symbols)\\
	&	Performance target:\\
	&	\,\,\,- Format 1: 1\% D2A and Aerr, 0.1\% N2A\\
	&	\,\,\,- Format 3: BLER 1\%\\
			
			\bottomrule
	\end{tabular}}
	
	\label{PUCCH_LS}
\end{table}

\begin{figure}[!t]
	\centering
	\includegraphics[width=\columnwidth]{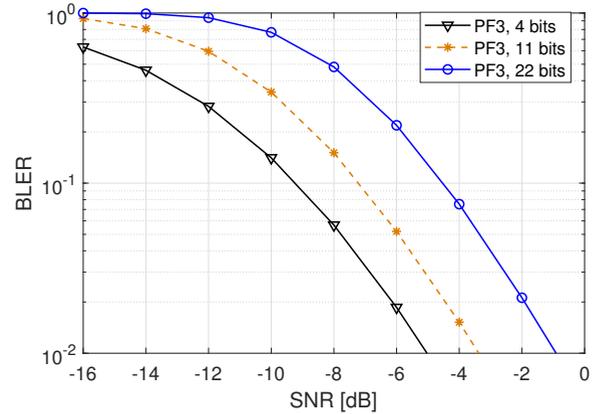}
	\caption{BLER performance of PUCCH format 3, 700 MHz.} 
	\label{PUCCH_700F3}
\end{figure}

\begin{figure}[!t]
	\centering
	\includegraphics[width=\columnwidth]{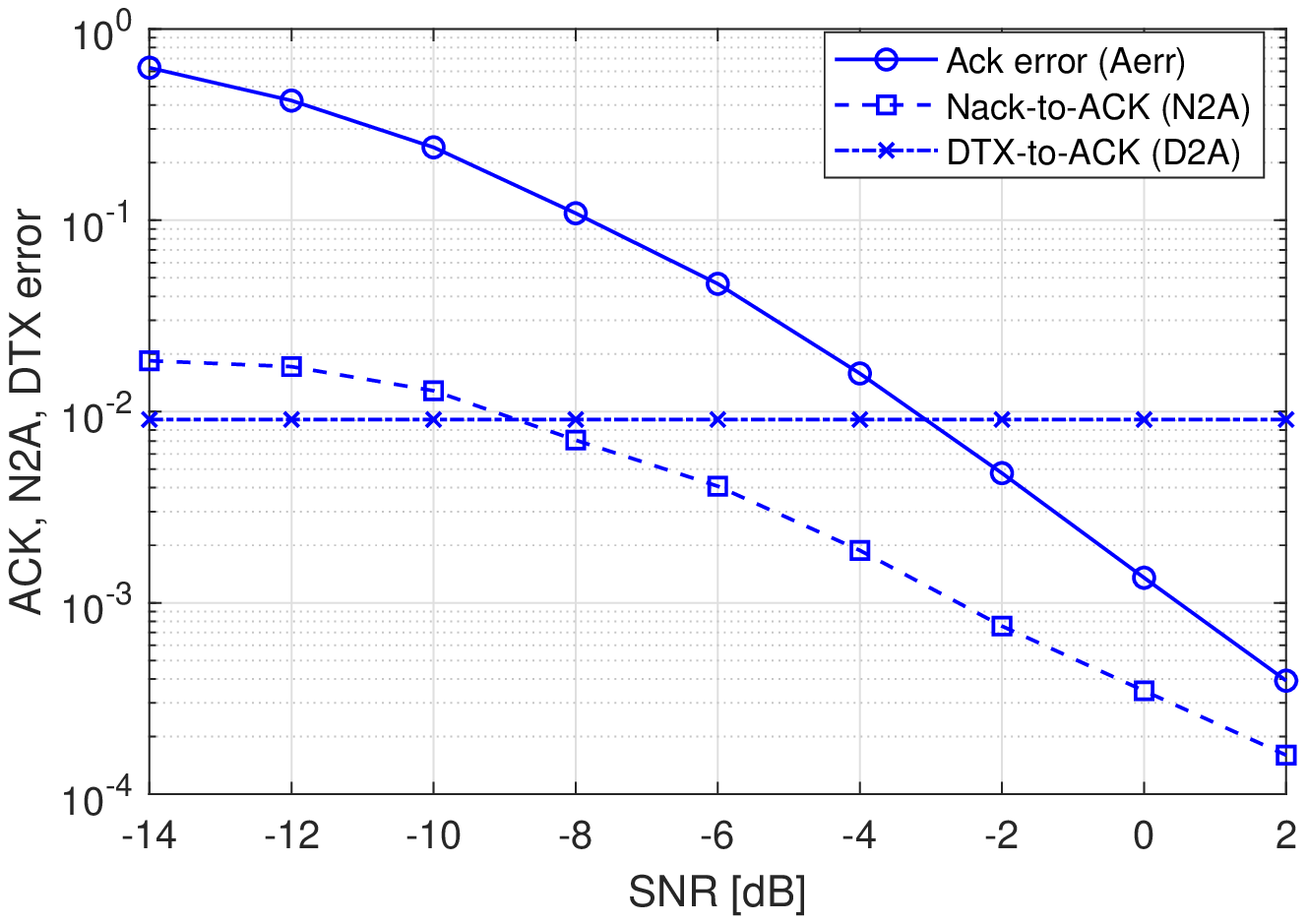}
	\caption{BLER performance of PUCCH format 1, 700 MHz.} 
	\label{PUCCH_700F1}
\end{figure}

\begin{figure}[!t]
	\centering
	\includegraphics[width=\columnwidth]{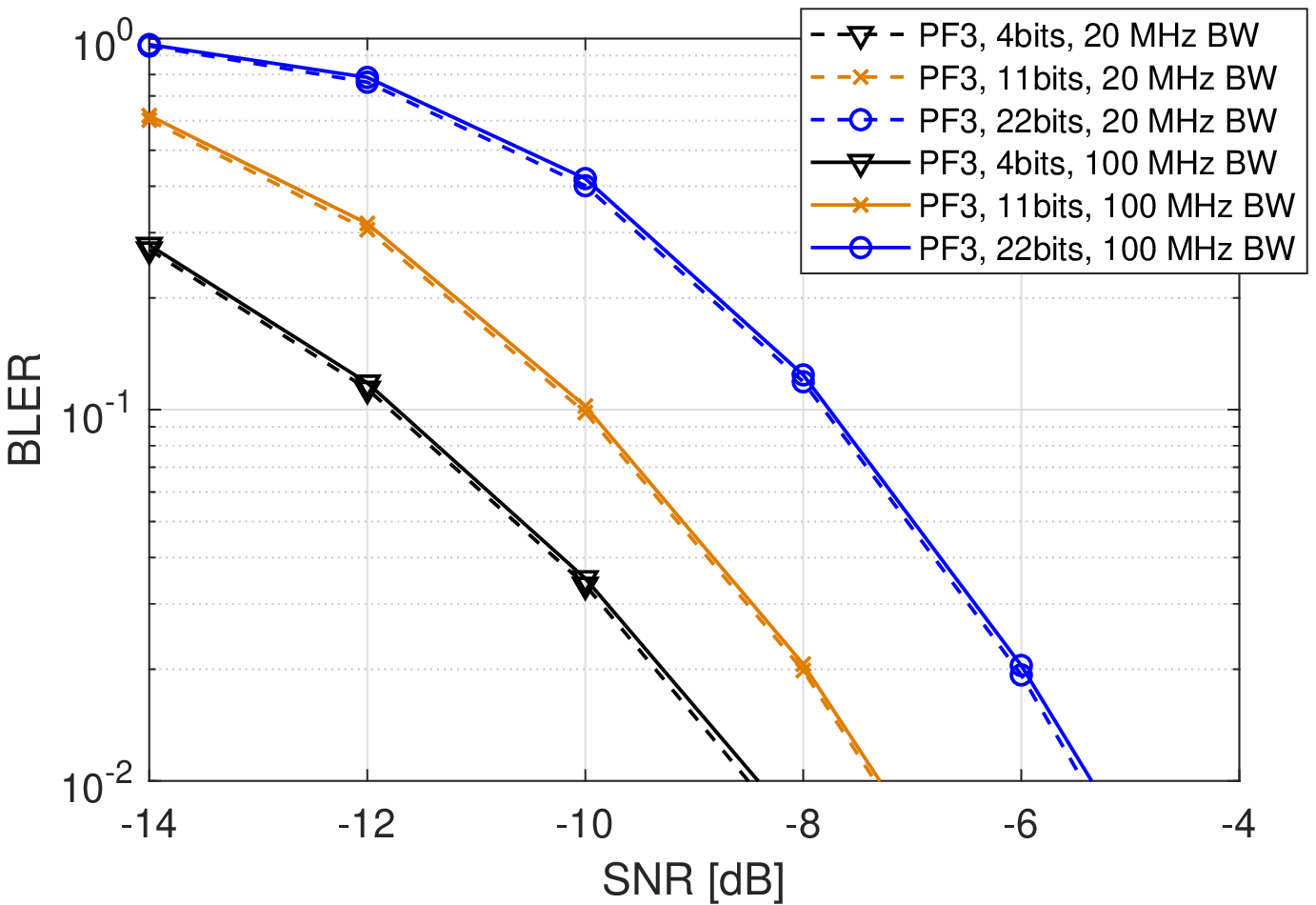}
	\caption{BLER performance of PUCCH format 3, 2.6 GHz.} 
	\label{PUCCH_2.6F3}
\end{figure}

\begin{figure}[!t]
	\centering
	\includegraphics[width=\columnwidth]{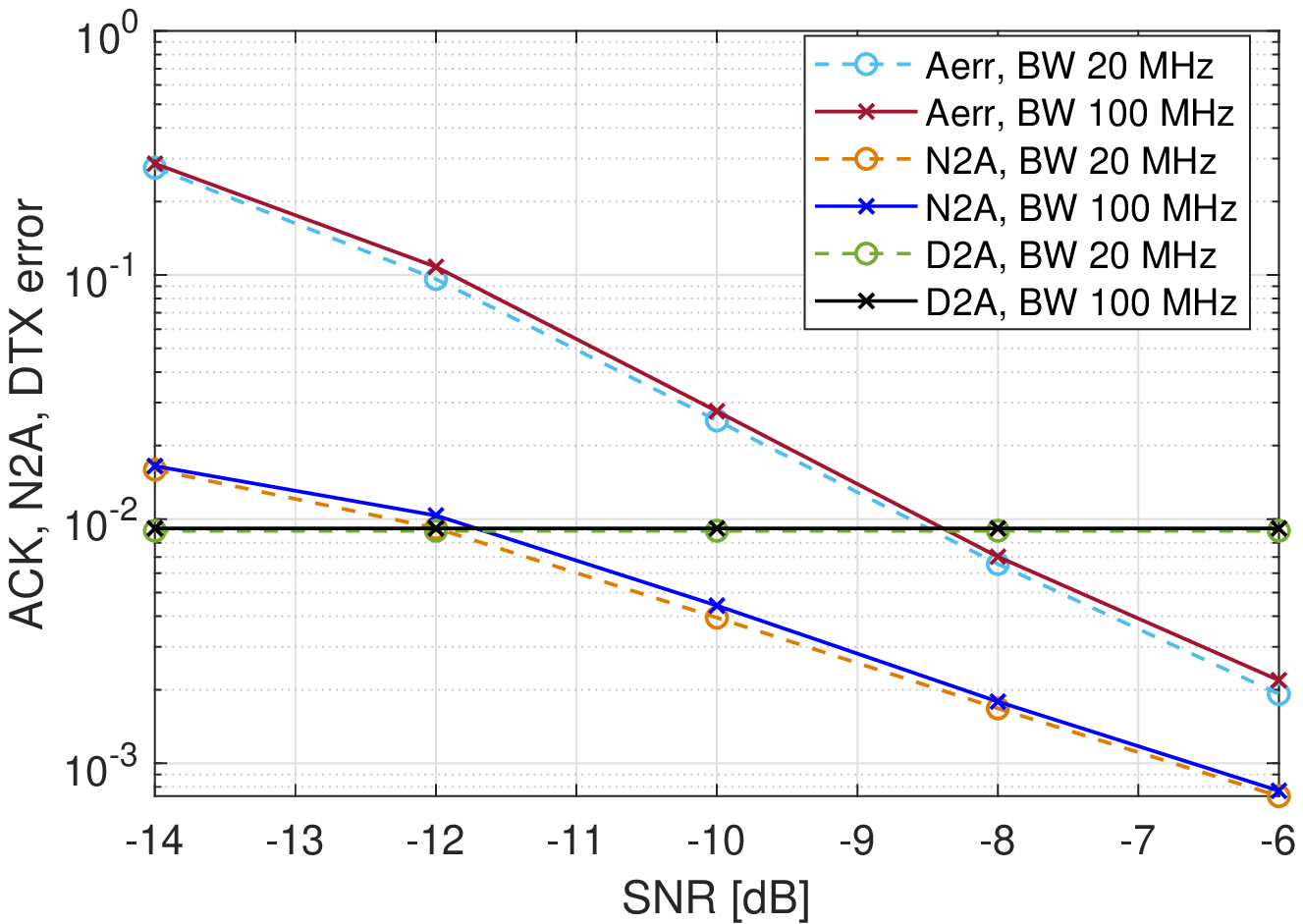}
	\caption{BLER performance of PUCCH format 1, 2.6 GHz.} 
	\label{PUCCH_2.6F1}
\end{figure}

\begin{figure}[!t]
	\centering
	\includegraphics[width=\columnwidth]{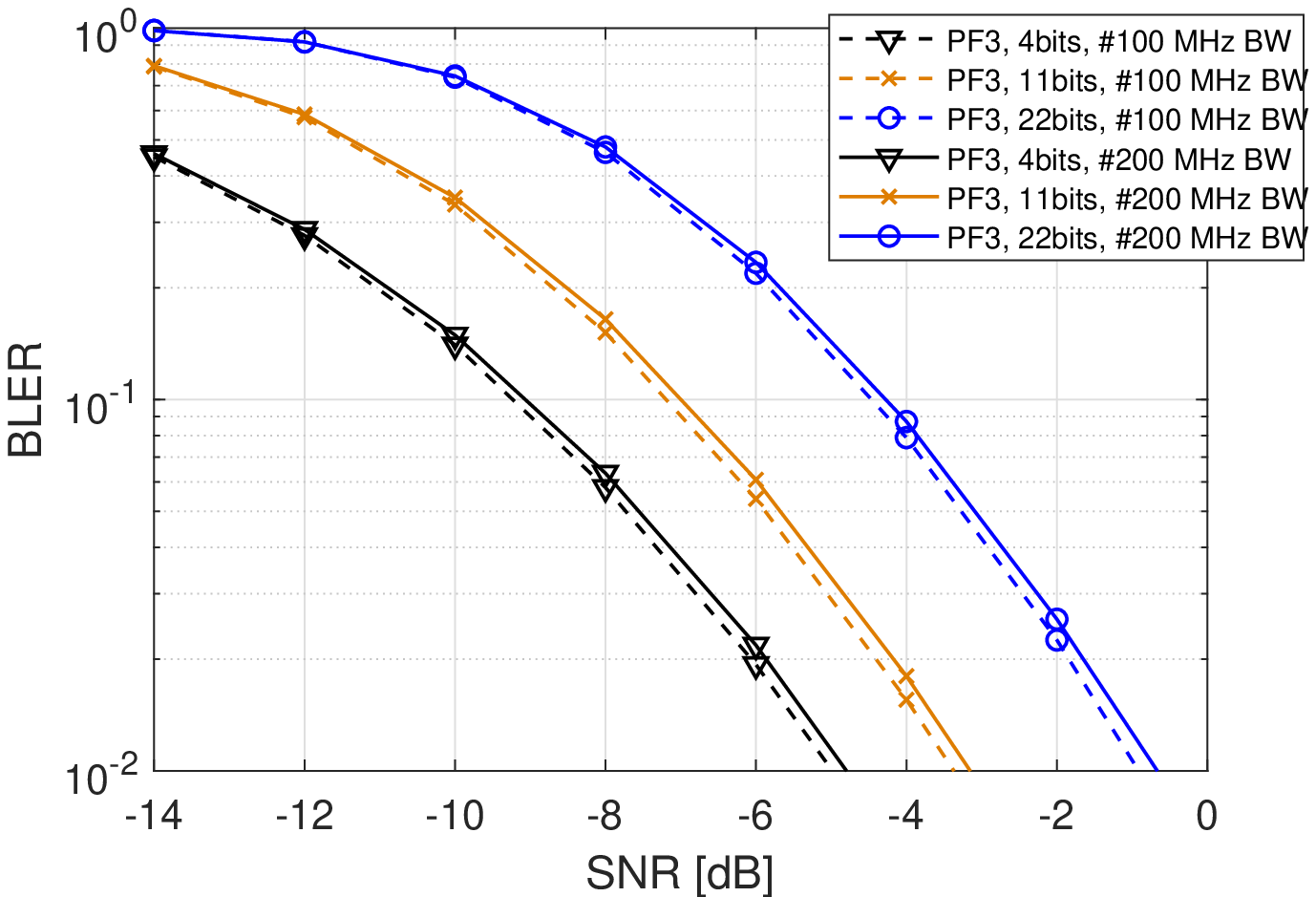}
	\caption{BLER performance of PUCCH format 3, 28 GHz.} 
	\label{PUCCH_28F3}
\end{figure}

\begin{figure}[!t]
	\centering
	\includegraphics[width=\columnwidth]{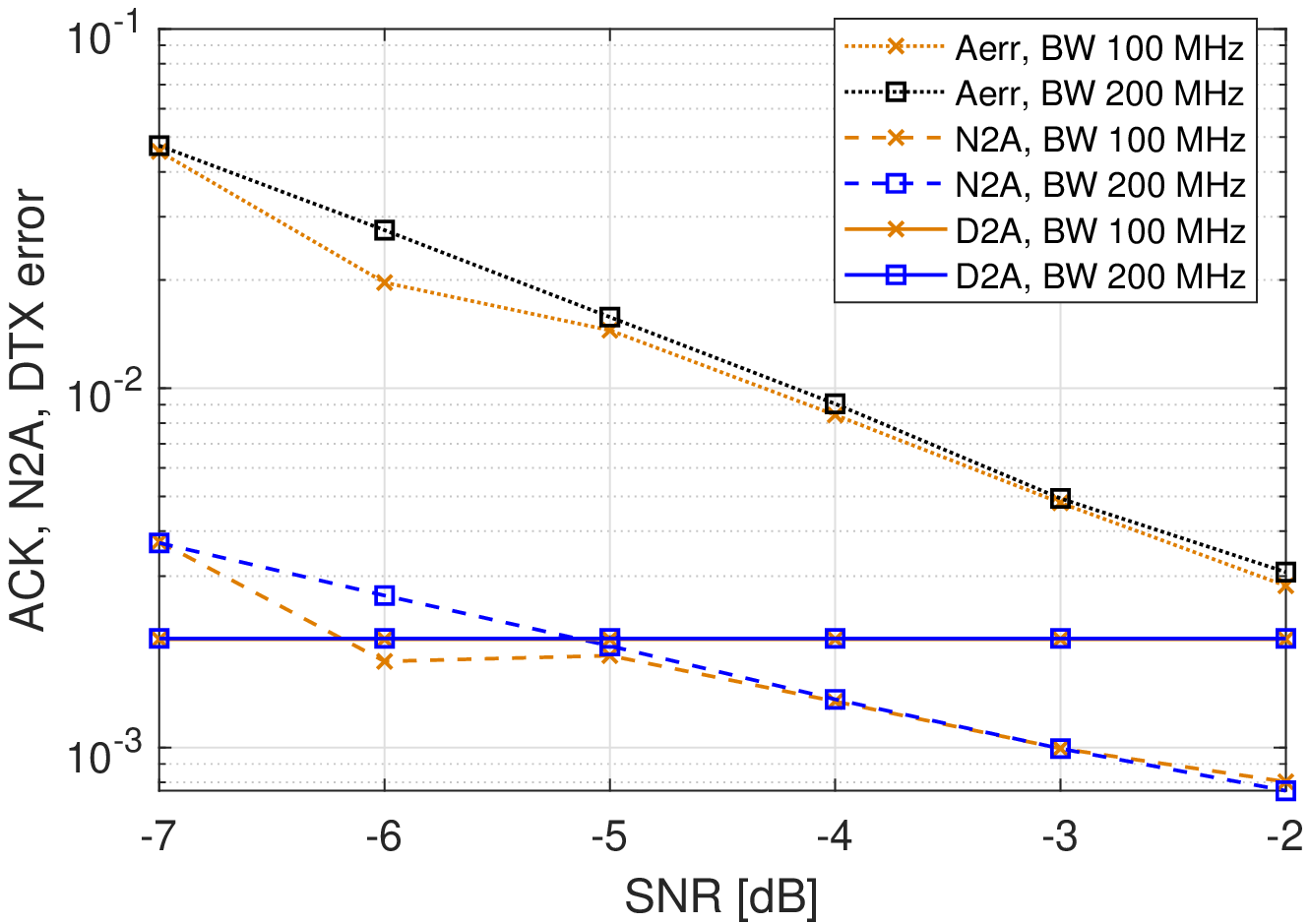}
	\caption{BLER performance of PUCCH format 1, 28 GHz.} 
	\label{PUCCH_28F1}
\end{figure}

Table \ref{PUCCH_LS} shows the channel-specific parameters and performance targets for PUCCH. The LLS results for PUCCH at carrier frequencies of 700 MHz, 2.6 GHz and 28 GHz are shown in Figures \ref{PUCCH_700F3}-\ref{PUCCH_28F1}. The results show that there is no significant performance impact due to complexity reduction in terms of reduced BW as the PUCCH frequency resource spans only 1 PRB. Since a single UE transmit antenna is assumed in the simulation for both RedCap and NR reference UE, there is no performance impact related to the reduction of the number of UE antennas.

\subsection{PUSCH}

\begin{table}[!t]
	
	\caption{\small Channel-specific parameters for PUSCH.}
	\resizebox{\columnwidth}{!}{
		\begin{tabular}{@{}ll@{}}
			\toprule
			Channel &	Assumptions \\ \hline
		PUSCH &	FDRA: \\
	 &	\,\,\,- Urban: 30 PRBs \\
	 &	\,\,\,- Indoor: 66 PRBs\\
	 &\,\,\,- 	Rural: 4 PRBs\\
	 &	TDRA: 14 OFDM symbols\\
	 &	Waveform: DFT-s-OFDM\\
	 &	DMRS: Type I, 2 DMRS symbol, no multiplexing with data \\
	 &	Target data rate/TBS/MCS: using MCS Table 6.1.4.1-2 (TS38.214 \cite{})  \\
	 &\,\,\,- Urban: 1 Mbps/TBS =552/MCS3\\
	 &\,\,\,- Indoor: 5 Mbps/TBS = 736/MCS1\\
	 &\,\,\,- Rural: 100 kbps/TBS = 128/MCS6\\
	 &	Rx combining: MRC\\
	 &	No frequency hopping\\
	 &	BLER target: 10\% \\
		
	\bottomrule
	\end{tabular}}
	
	\label{PUSCH_LS}
\end{table}

\begin{figure}[!t]
	\centering
	\includegraphics[width=\columnwidth]{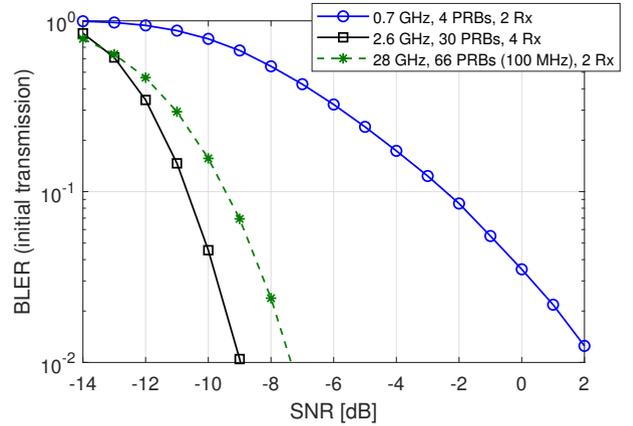}
	\caption{BLER performance of PUSCH.} 
	\label{PUSCH_BLER}
\end{figure}

\begin{figure}[!t]
	\centering
	\includegraphics[width=\columnwidth]{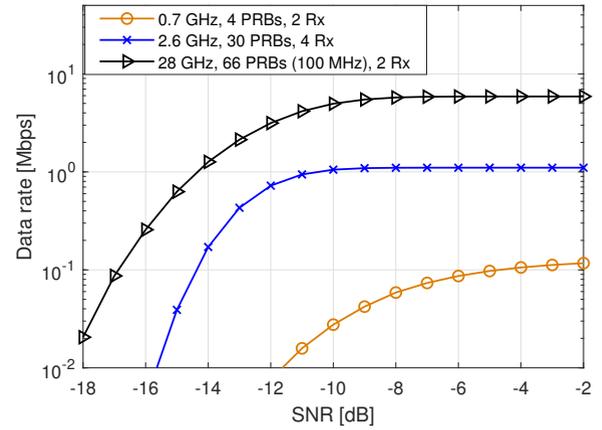}
	\caption{Data rate for PUSCH.} 
	\label{PUSCH_rate}
\end{figure}

Our assumptions for performing PUSCH LLSs are shown in Table \ref{PUSCH_LS}. Figure \ref{PUSCH_BLER} and Figure \ref{PUSCH_rate} show the BLER performance and data rate of the PUSCH for different carrier frequencies. Similar to other uplink physical channels considered in this paper, the number of Tx branches is the same at the reference UE and the RedCap UE. Furthermore, as shown in Table \ref{PUSCH_LS}, the PUSCH transmission BW is assumed to be less than that of the RedCap UE BW in Urban, Indoor and Rural scenarios. Therefore, the link performance will be identical for the RedCap UE and the reference UE.

\section{LINK BUDGET EVALUATION}

Link budget evaluation is used to investigate coverage by tracking the transmitted power, the gains and the losses along the transmission path and power is sufficient so that the system can operate acceptably. Coverage can be expressed by different metrics such as maximum coupling loss (MCL), maximum path loss (MPL) and maximum isotropic loss (MIL) \cite{TR38830}. Among these metrics, MIL and MPL include the antenna gains. However, compared to MPL, MIL is more straightforward to compute  as it does not consider parameters such as shadow fading and penetration margins. Therefore, in this paper, we have used MIL as the key coverage evaluation metric.
Considering the simulation results and the corresponding performance targets for the different physical channels, the required SINRs to fulfill these targets are reported in  Tables \ref{SNR_700}-\ref{SNR_28}, respectively, for Rural, Urban, and Indoor scenarios.

\begin{table}[!t]
	\centering 
	\caption{\small Required SINR (dB), 700 MHz.}
	\resizebox{.8\columnwidth}{!}{
		\begin{tabular}{ l c c } 
			
			\hline
			
			&  BW= 20 MHz,  2Rx &	BW= 20 MHz,  1Rx	\\\hline
			SSB&	-7.3&	-2.9\\\hline
			PRACH&	-9.2&	-9.2\\\hline
			Msg2&	-9.6&	-5.8\\\hline
			Msg3&	-1.5&	-1.5\\\hline
			Msg4&	-5.9&	-2.2\\\hline
			PDCCH&	-6.6&	-3.1\\\hline
			PDSCH&	-5.6&	-2\\\hline
			PUCCH (2 bits)& 	-2.9&	-2.9\\\hline
			PUCCH (11 bits)& 	-3.4&	-3.4\\\hline
			PUCCH (22 bits)&	-0.9&	-0.9\\\hline
			PUSCH&	-2.4&	-2.4\\\hline

	\end{tabular}}
	
	\label{SNR_700}
	
\end{table}

\begin{table}[!t]
	\centering 
	\caption{\small Required SINR (dB), 2.6 GHz.}
	\resizebox{1\columnwidth}{!}{
		\begin{tabular}{ l c c c } 
			\hline
	&BW= 100 MHz,  4Rx &	BW= 20 MHz, 2Rx
	&BW= 20 MHz, 1Rx	\\\hline
SSB	& -11 &	-8	&-4.1 \\\hline
PRACH &	-17.5&	-17.5&	-17.5 \\\hline
Msg2&	-9.6&	-6.5&	-3.1 \\\hline
Msg3&	-6&	-6&	-6 \\\hline
Msg4&	-6.6&	-3.1&	0.9 \\\hline
PDCCH&	-9.2&	-6&	-3 \\\hline
PDSCH&	-5.7&	-2.7&	0.5 \\\hline
PUCCH (2 bits)& 	-6.6&	-6.6&	-6.6 \\\hline
PUCCH (11 bits)& 	-7.3&	-7.3&	-7.3 \\\hline
PUCCH (22 bits)&	-5.3&	-5.3&	-5.3 \\\hline
PUSCH&	-10.5 &	-10.5&	-10.5 \\\hline
	\end{tabular}}
	
	\label{SNR_2.6}

\end{table}

\begin{table}[!t]
	\centering 
	\caption{\small Required SINR (dB), 28 GHz.}
	\resizebox{.8\columnwidth}{!}{
		\begin{tabular}{ l c c } 
			\hline
			&BW= 100 MHz,  2Rx &	BW= 100 MHz,  1Rx	\\\hline
		SSB &	-8.2 &	-4.5 \\\hline
		PRACH&	-12.2&	-12.2 \\\hline
		Msg2&	-9.4&	-6 \\\hline
		Msg3&	-1.8&	-1.8 \\\hline
		Msg4&	-5.4&	-1.4 \\\hline
		PDCCH&	-6&	-2.1 \\\hline
		PDSCH&	-2.3&	1.7 \\\hline
		PUCCH (2 bits)& 	-3.02&	-3.02 \\\hline
		PUCCH (11 bits)& 	-3.37&	-3.37 \\\hline
		PUCCH (22 bits)&	-0.91&	-0.91 \\\hline
		PUSCH&	-9.4&	-9.4 \\\hline
		
	\end{tabular}}
	
	\label{SNR_28}
	
\end{table}

The SINR values shown in these tables are used to perform link budget evaluation based on the template \cite{E_diss}. Table \ref{Link_bud} shows the key assumptions that we have considered in our link budget evaluations. It should be noted  that for RedCap UEs operating in FR1 band (Rural and Urban), due to device size limitations, we have considered additional 3 dB antenna inefficiency compared to the reference NR UEs.

\begin{table}[!t]
	
	\caption{\small Link budget assumptions.}
	\resizebox{0.9\columnwidth}{!}{
		\begin{tabular}{@{}ll@{}}
			\toprule
	Parameter name &	Value\\ \hline
	gNB total transmit power 
	for carrier bandwidth (dBM)	 &Rural: 49\\
 &	Urban: 53\\
 &	Indoor: 23\\ \hline
	Number of gNB  TXRUs &	Rural: 2 \\
&	Urban: 64\\
&	Indoor: 2\\ \hline
	UE total transmit power (dBm)
	for carrier bandwidth (dBM) &	Rural: 23\\
&	Urban: 23\\
&	Indoor: 12\\		
	\bottomrule
	\end{tabular}}
	
	\label{Link_bud}
\end{table}

\begin{figure}[!t]
	\centering
	\includegraphics[width=\columnwidth]{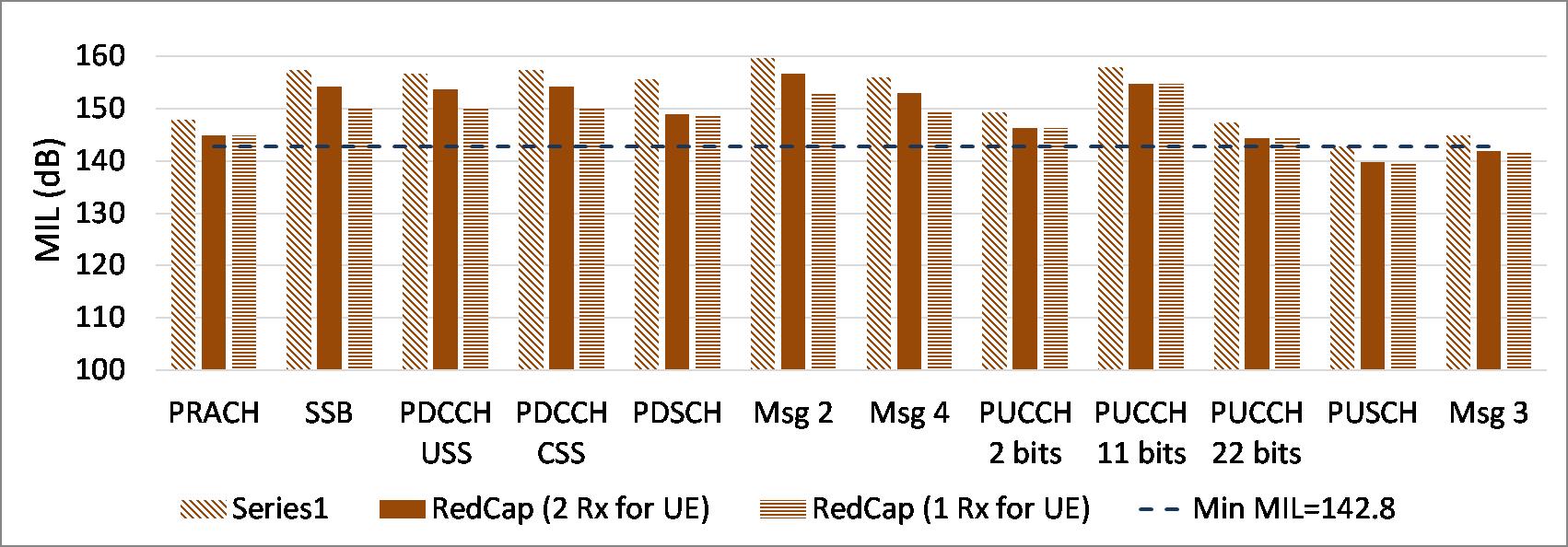}
	\caption{MIL for 700 MHz.} 
	\label{MIL_700}
\end{figure}

\begin{figure}[!t]
	\centering
	\includegraphics[width=\columnwidth]{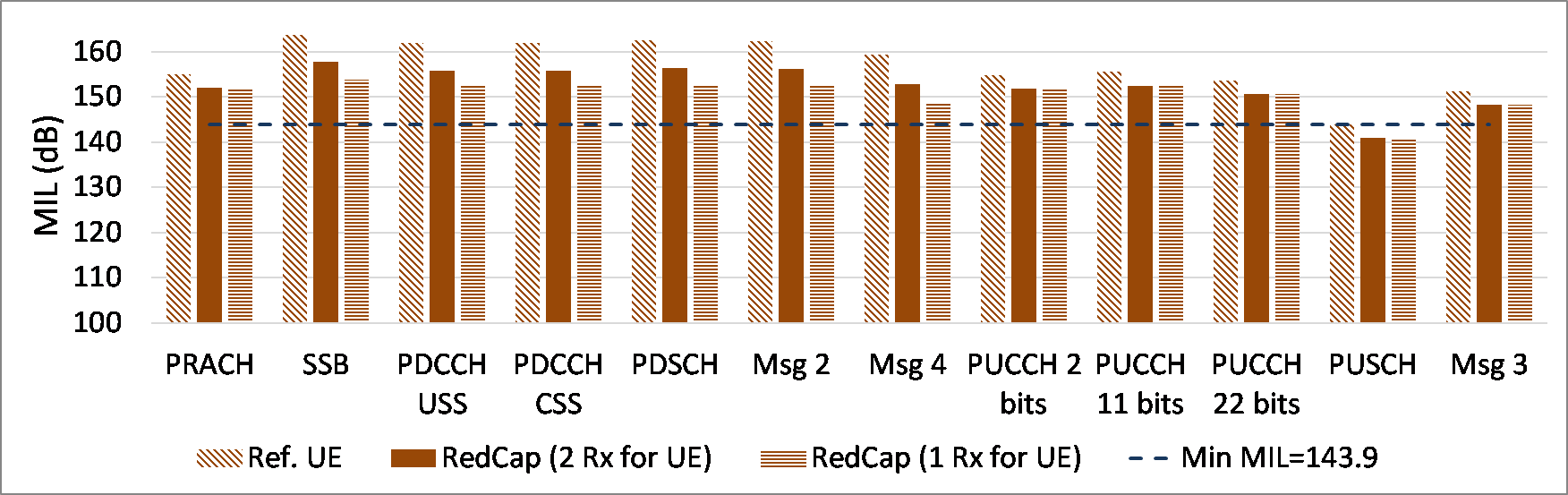}
	\caption{MIL for 2.6 GHz.} 
	\label{MIL_2.6}
\end{figure}

\begin{figure}[!t]
	\centering
	\includegraphics[width=\columnwidth]{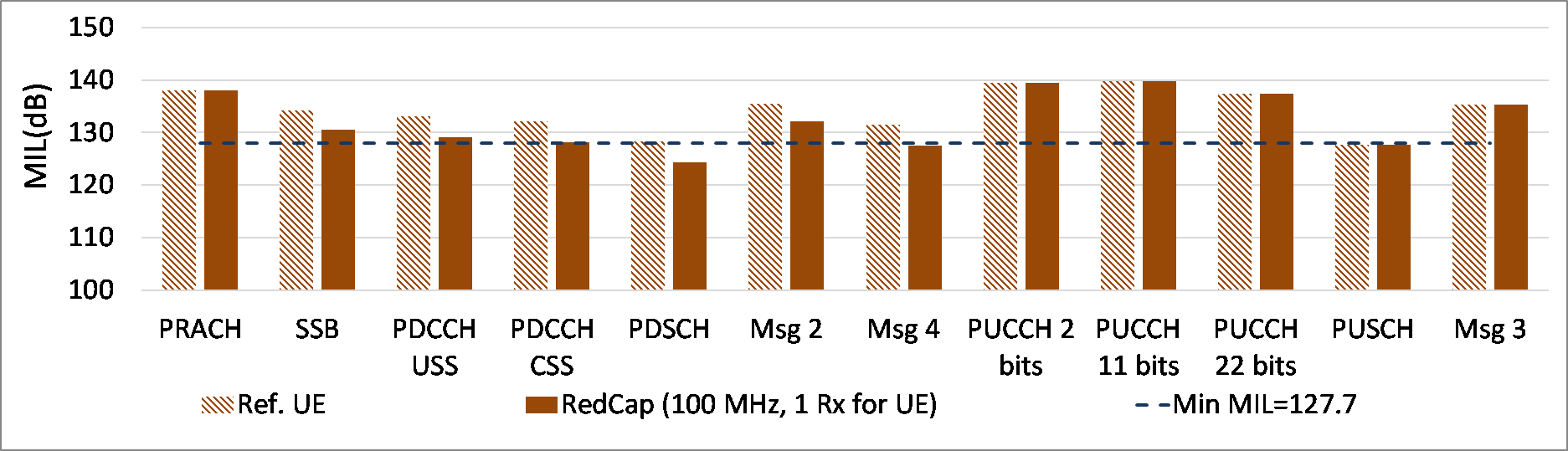}
	\caption{MIL for 28 GHz.} 
	\label{MIL_28}
\end{figure}

In Figures \ref{MIL_700}-\ref{MIL_28} the coverage of different RedCap physical channels in terms of MIL are compared to that of the corresponding NR channels at carrier frequencies of 700 MHz 2.6 GHz and 28 GHz, respectively. At each of the scenarios, the NR physical channel with the lowest MIL is considered as coverage bottleneck channel, i.e. the corresponding value is the minimum acceptable MIL and the Rel-15 NR coverage limit is assumed to be given by this MIL. We have considered this MIL value as the minimum acceptable MIL also for RedCap channels and used it as a threshold to identify the RedCap physical channels that need coverage recovery. Any RedCap channel whose MIL is worse than that of the threshold MIL needs coverage recovery and the amount of required coverage compensation is the difference of the RedCap-channel-MIL and the threshold MIL.

As it can be seen in Figure \ref{MIL_700} and Figure \ref{MIL_2.6}, for the reference
UE operating in Rural and Urban scenarios, PUSCH
is the bottleneck channel and has the lowest MIL value (MIL~
=~ 142.8 dB for Rural and MIL = 143.9 dB for Urban). For RedCap UEs operating in Rural scenario, all
the physical channels and initial access messages, except
PUSCH and Msg3, have MIL larger than the threshold value. Based on our results for Rural case, PUSCH and Msg3 need 3 dB and 0.8 dB coverage compensation.
For RedCap UEs operating in Urban scenarios, only PUSCH needs 3 dB coverage compensation.

For the reference UE in Indoor scenario, as it is shown in Figure \ref{MIL_28}, PUSCH is the bottleneck channel with MIL~=~127.7 dB. For RedCap UE with 1 Rx branch, coverage compensations of approximately 3.4 dB and 0.5 dB are respectively, needed for PDSCH and Msg4.

\section{CONCLUSIONS}
In this paper, we have investigated the coverage performance of the NR-RedCap UEs and identified the physical channels that limit the coverage of these devices. We first have provided an overview of the NR-RedCap and discussed its use cases, requirements, and main features. Then, for different deployment scenarios and carrier frequencies (FR1 and FR2), we have evaluated the link performance of RedCap UEs and performed link-budget evaluations for all physical channels and messages for DL and UL transmissions.

Our results have shown that for RedCap UEs operating in FR1 band, the PUSCH  can limit the coverage, and it needs 3 dB coverage recovery. It is worth to highlight two observations; first, the 3 dB coverage loss resulting from the UE antenna efficiency loss due to device size limitations; second, by reducing the data rate target for RedCap UEs in UL, no coverage recovery is needed. For the Rural case, a small amount of coverage compensation (approximately 0.8 dB) is needed for Msg3. For RedCap UEs operating in FR2 band, the impact of complexity reduction is more considerable for DL channels, and PDSCH and Msg4 are the channels that may need coverage recovery. However, the amount of coverage-compensation needed for Msg4 is less than 0.5 dB and by considering smaller data rates no coverage recovery is needed for PDSCH.

\bibliographystyle{IEEEtran} 
\bibliography{IEEEabrv, reference}

\begin{thebibliography}{10}
\providecommand{\url}[1]{#1}
\csname url@samestyle\endcsname
\providecommand{\newblock}{\relax}
\providecommand{\bibinfo}[2]{#2}
\providecommand{\BIBentrySTDinterwordspacing}{\spaceskip=0pt\relax}
\providecommand{\BIBentryALTinterwordstretchfactor}{4}
\providecommand{\BIBentryALTinterwordspacing}{\spaceskip=\fontdimen2\font plus
\BIBentryALTinterwordstretchfactor\fontdimen3\font minus
  \fontdimen4\font\relax}
\providecommand{\BIBforeignlanguage}[2]{{%
\expandafter\ifx\csname l@#1\endcsname\relax
\typeout{** WARNING: IEEEtran.bst: No hyphenation pattern has been}%
\typeout{** loaded for the language `#1'. Using the pattern for}%
\typeout{** the default language instead.}%
\else
\language=\csname l@#1\endcsname
\fi
#2}}
\providecommand{\BIBdecl}{\relax}
\BIBdecl

\bibitem{dahlman20185g}
E.~Dahlman, S.~Parkvall, and J.~Skold, \emph{{5G NR}: The next generation
  wireless access technology}.\hskip 1em plus 0.5em minus 0.4em\relax Academic
  Press, 2018.

\bibitem{ahmadi}
S.~Ahmadi, \emph{{5G NR}: Architecture, Technology, Implementation, and
  Operation of {3GPP} New Radio Standards}.\hskip 1em plus 0.5em minus
  0.4em\relax Academic Press, 2019.

\bibitem{GSMAwhite}
{GSMA}, ``Mobile {IoT} in the {5G} future-{NB-IoT} and {LTE-M} in the context
  of {5G},'' \emph{White paper}, 2018.

\bibitem{liberg2017cellular}
O.~Liberg, M.~Sundberg, E.~Wang, J.~Bergman, and J.~Sachs, \emph{Cellular
  {Internet of Things}: Technologies, Standards, and Performance}.\hskip 1em
  plus 0.5em minus 0.4em\relax Academic Press, 2017.

\bibitem{NRCoex}
M.~{Mozaffari}, Y.-P.~E. {Wang}, O.~{Liberg}, and J.~{Bergman}, ``Flexible and
  efficient deployment of {NB-IoT} and {LTE-MTC} in coexistence with {5G} new
  radio,'' in \emph{Proc. IEEE INFOCOM}, April 2019, pp. 391--396.

\bibitem{SID_redcap}
{3GPP Tdoc RP-201677, Ericsson}, ``Revised {SID} on study on support of reduced
  capability {NR} devices,'' July 2020.

\bibitem{WID_redcap}
{3GPP Tdoc RP-202933, Ericsson and Nokia}, ``New {WID} on support of reduced
  capability {NR} devices,'' Dec. 2020.

\bibitem{TR38875}
{3GPP TR 38.875}, ``Study on support of reduced capability {NR} devices
  ({Release 17}),'' Dec. 2020.

\bibitem{TS38214}
{3GPP TS 38.214}, ``{NR}; physical layer procedures for data ({Release 16}),''
  2019.

\bibitem{TR38830}
{3GPP TR 38.830}, ``Study on {NR} coverage enhancements ({Release 17}),'' Dec.
  2020.

\bibitem{E_diss}
{ 3GPP Tdoc R1-2007481, Moderator (Ericsson, Apple, Qualcomm)}, ``Feature lead
  ({FL}) summary 4 for redcap evaluation templates, {RAN1} 102-e,'' Aug. 2020.

\end{thebibliography}

\EOD

\end{document}